\documentclass[longauth]{aa}

\usepackage{graphicx}

\usepackage{txfonts}

\usepackage{xcolor}
\usepackage{placeins}

\begin{document} 

\title{The ALMA survey to Resolve exoKuiper belt Substructures (ARKS)}

\subtitle{XII. Constraints on large bodies embedded in discs}

\titlerunning{Constraints on large bodies embedded in discs}

\author{M.~R.~Jankovic\textsuperscript{1}\fnmsep\thanks{E-mail: marija.jankovic@ipb.ac.rs} \and A.~V.~Krivov\textsuperscript{2} \and T.~D.~Pearce\textsuperscript{3,4} \and A.~A.~Sefilian\textsuperscript{5} \and M.~C.~Wyatt\textsuperscript{6} \and E.~Chiang\textsuperscript{7} \and C.~del~Burgo\textsuperscript{8,9} \and Y.~Han\textsuperscript{10} \and J.~Jovicic\textsuperscript{11} \and S.~Marino\textsuperscript{12} \and B.~Zawadzki\textsuperscript{13} \and A.~M.~Hughes\textsuperscript{13} \and T.~L\"ohne\textsuperscript{2} \and P.~Luppe\textsuperscript{3} \and J.~P.~Marshall\textsuperscript{14} \and Th.~Henning\textsuperscript{15} \and H.~E.~Schlichting\textsuperscript{16} \and S.~Mac~Manamon\textsuperscript{3} \and N.~Pawellek\textsuperscript{17,18} \and D.~J.~Wilner\textsuperscript{19}}

\authorrunning{Jankovic, M. R., et al.}

\institute{
Institute of Physics Belgrade, University of Belgrade, Pregrevica 118, 11080 Belgrade, Serbia \and
Astrophysikalisches Institut und Universit\"atssternwarte, Friedrich-Schiller-Universit\"at Jena, Schillerg\"a{\ss}chen 2-3, 07745 Jena, Germany \and
School of Physics, Trinity College Dublin, the University of Dublin, College Green, Dublin 2, Ireland \and
Department of Physics, University of Warwick, Gibbet Hill Road, Coventry CV4 7AL, UK \and
Department of Astronomy and Steward Observatory, The University of Arizona, 933 North Cherry Ave, Tucson, AZ, 85721, USA \and
Institute of Astronomy, University of Cambridge, Madingley Road, Cambridge CB3 0HA, UK \and
Department of Astronomy, University of California, Berkeley, Berkeley, CA 94720-3411, USA \and
Instituto de Astrof\'isica de Canarias, V\'ia L\'actea S/N, La Laguna, E-38200, Tenerife, Spain \and
Departamento de Astrof\'isica, Universidad de La Laguna, La Laguna, E-38200, Tenerife, Spain \and
Division of Geological and Planetary Sciences, California Institute of Technology, 1200 E. California Blvd., Pasadena, CA 91125, USA \and
Department of Astronomy, Faculty of Mathematics, University of Belgrade, Studentski trg 16, 11000 Belgrade, Serbia \and
Department of Physics and Astronomy, University of Exeter, Stocker Road, Exeter EX4 4QL, UK \and
Department of Astronomy, Van Vleck Observatory, Wesleyan University, 96 Foss Hill Dr., Middletown, CT, 06459, USA \and
Academia Sinica Institute of Astronomy and Astrophysics, 11F of AS/NTU Astronomy-Mathematics Building, No.1, Sect. 4, Roosevelt Rd, Taipei 106319, Taiwan. \and
Max-Planck-Insitut f\"ur Astronomie, K\"onigstuhl 17, 69117 Heidelberg, Germany \and
Department of Earth, Planetary, and Space Sciences, The University of California, Los Angeles, 595 Charles E. Young Drive East, Los Angeles, CA 90095, USA \and
Institut f\"ur Astrophysik, Universit\"at Wien, T\"urkenschanzstra\ss{}e 17, 1180 Wien, Austria \and
Konkoly Observatory, HUN-REN Research Centre for Astronomy and Earth Sciences, MTA Centre of Excellence, Konkoly-Thege Mikl\'os \'ut 15-17, 1121 Budapest, Hungary \and
Center for Astrophysics | Harvard \& Smithsonian, 60 Garden St, Cambridge, MA 02138, USA
}

\date{Received MM DD, YEAR; accepted MM DD, YEAR}

\abstract
{The ALMA survey to Resolve exoKuiper belt Substructures (ARKS) presents detailed constraints on structure and properties of a sample of 24 debris discs at millimetre wavelengths, including measurements of vertical thicknesses for a subsample of discs.}
{We estimated sizes and total masses of large bodies that feed collisional cascades producing the observed millimetre dust in 11 ARKS discs with a well constrained vertical thickness. Separately, we estimated the sizes and total masses of large bodies that could be gravitationally stirring these discs, assuming that such stirring is the origin of the vertical thickness.}
{We used a semi-analytic particle-in-a-box collisional model and a semi-analytic model of stirring through gravitational scattering. We accounted for the dependence of collisional damping efficiency on the average collision velocity.}
{We find that planetesimals feeding the collisional cascades are less than 20\,km in size, while the total masses in the planetesimals are lower than 200\,M$_\oplus$. There is a good agreement between minimum disc masses inferred from millimetre and infrared observations. The hypothetical stirrers are always larger in size than the planetesimals, but they do not necessarily contain most of the disc mass.}
{In agreement with previous work, the observed dust masses can be explained by steady-state collisional models with reasonable disc masses. The measured vertical scale-heights can be explained by gravitational stirring from embedded large bodies of reasonable properties. However, more advanced models are needed to address the observed non-Gaussian vertical profiles, along with the dependence of disc scale-height on particle size and stellocentric distance.}


\maketitle

\section{Introduction}
Cold dusty discs surround many main sequence stars. The dust is believed to be continually produced in large debris collisions that make up the bulk of the discs' masses, but that cannot be directly detected \citep{Wyatt2008, Krivov2010, Matthews2014, Hughes2018, Wyatt2020, Marino2022_review, Pearce2026_review}. These large bodies are analogous to asteroids, Kuiper belt objects, and comets in the Solar System, and the small dust is key to our understanding of these bodies around other stars.

Using models of collisional cascades and measurements of dusty disc properties, it is possible to estimate the sizes and the total masses of the large bodies whose collisions produce the dust. We refer to these bodies as planetesimals. Current models predict that the size distribution of particles in a debris collisional cascade approximately follows a broken power law \citep[e.g.][]{O'Brien2003,Pan2005,Pan2012,Thebault2003,Krivov2006}. The observed dust mass can be extrapolated using the power-law size distribution to find the mass contained in the cascade as a function of the size of the largest bodies in the cascade. The size of the largest bodies in the cascade can be estimated as the size at which the collisional fragmentation timescale equals the system age. This timescale can be estimated through a simple particle-in-a-box model \citep{Wyatt2002}. There might be bodies in the disc that are larger than those in the cascade, but they have longer fragmentation timescales and typically they have not yet collided. Smaller bodies, on the other hand, fragment and get removed from the system by radiation forces and/or a stellar wind on timescales shorter than the system age. In turn, they need to be replenished by larger bodies for the cascade to remain observable. With this approach, the so-called collisional mass of a debris disc can be estimated \citep[e.g.][]{Quillen2007,Krivov2021,Pearce2025}.

Furthermore, dusty discs may also contain (dwarf) planets that gravitationally stir and/or shape the discs. From resolved debris disc structures, dynamical constraints can be placed on the size and the number of these large bodies. Broadly speaking, there are two main system architectures that are typically explored. It can be assumed that large bodies within the disc viscously stir planetesimals and the dust \citep{Ida1993}. Such bodies could either have formed in the protoplanetary disc phase \citep{Krivov2018b}, or their accretion may be ongoing throughout the debris disc evolution \citep{Kenyon2002b, Kenyon2004a, Kenyon2008, Kenyon2010}. In the former case, simple semi-analytic models can be used to relate disc observables to the sizes and the total mass of the stirrers \citep[e.g.][]{Quillen2007,Daley2019,Matra2019,Terrill2023}. On the other hand, the bodies stirring and shaping debris discs could be outside the discs, affecting discs via scattering as well as secular and resonant interactions. Such external perturbers could be closer to the star than the disc \citep[e.g.][]{Costa2024,Pearce2024,Sefilian2024}, farther away \citep[e.g.][]{Nesvold2017,Stuber2023}, or both \citep{Boley2012,Farhat2023}. In this work, we focus on the former set-up, with hypothetical stirrers embedded in discs.

Recently, the Atacama Large Millimeter/submillimeter Array (ALMA) survey to Resolve exoKuiper belt Substructures \citep[ARKS;][]{Marino2026a} measured the structure of 24 debris discs in unprecedented detail at millimetre wavelengths. In the paper dedicated to the characterisation of the disc's vertical structures, \citet{Zawadzki2026} presented an analysis of a subsample of particularly well-resolved ARKS discs, inferring the sizes of bodies and disc masses required to stir the ARKS discs to their present-day vertical thicknesses. Based on assumptions of a balance between stirring and efficient collisional damping, they found the total mass of stirrers to range between 0.83 and 114\,M$_\oplus$. However, for some of the discs, the corresponding stirring timescale was found a posteriori to be longer than the estimated system age. Additionally, for some of the discs, these dynamical masses are smaller than the minimum masses previously derived \citep[albeit the latter have been estimated using different observational data, and also not for all of the ARKS discs;][]{Krivov2021}.

In this paper, we expand upon both the collisional and the dynamical constraints on large bodies hiding in ARKS discs. We consider these constraints independently, assuming that the planetesimals feeding the collisional cascades and the stirrers may be two distinct populations of large bodies. 
In Sect. \ref{sec:data}, we describe the dataset characterising our chosen sample of discs. In Sect. \ref{sec:parent_bodies}, we present the constraints placed on bodies in the collisional cascades. In Sect. \ref{sec:stirrers}, we revisit the constraints placed on the large bodies stirring the discs. In Sect. \ref{sec:discussion}, we bring together the two sets of constraints and discuss our results. In Sect. \ref{sec:conclusions}, we present our conclusions.

\section{Data} \label{sec:data}
We considered a sample of 11 discs with well-constrained vertical structure, originally presented in \citet{Zawadzki2026}. The data we used, which are summarised in Table \ref{table:data}, comprise a subsample of 24 discs analysed in the (sub)millimetre in the ARKS programme \citep{Marino2026a}. We adopted stellar masses ($M_*$), stellar luminosities ($L_*$), stellar ages ($t_{\rm age}$), and fractional luminosities ($f$) from Table 1 of \citet{Marino2026a}. In their work, \citet{Marino2026a} fit the discs' observations with a simple parametric Gaussian model of dust surface density and constant opacity. From Table B.1 in their work, we adopted the stellocentric radii of peak density ($r$, centre of the fitted Gaussian) and disc widths ($\Delta r$, FWHM of the fitted Gaussian), and the dust masses ($M_{\rm dust}$). The dust masses are derived by \citet{Marino2026a} assuming a power-law size distribution with a slope of $q=3.5$, with grains from 1\,\textmu m up to 1\,cm contributing to the observed flux. In this work, we assumed a slightly steeper size distribution at these dust sizes (see Sect. \ref{sec:size_distr}). We checked that changing the slope in the derivation of the dust mass to our theoretical one would only change the dust mass by 10-15\,\%; thus, we opted to neglect this small discrepancy.

For the disc aspect ratios ($h$), we adopted values from Table 3 of \citet{Zawadzki2026}. We only considered the 11 discs whose aspect ratios had been well constrained with parametric modelling to keep our analysis uniform. The modelling defines the disc scale-height as the half width at half maximum of the vertical density profile, and this scale-height is assumed to be directly proportional to stellocentric distance  within each disc \citep[see Eq. 1 in][]{Zawadzki2026}. The aspect ratio, defined as the ratio of the scale-height and stellocentric distance, is therefore assumed to be constant and does not depend on the radius within the disc.

\section{Constraints on planetesimals} \label{sec:parent_bodies}
In this section, we interpret the discs in our sample as steady-state collisional cascades. We assume that the observed disc aspect ratios determine collision velocities and that the velocities have been constant throughout the system lifetimes. Accounting for velocity evolution due to stirring by embedded large bodies would not significantly change our results (see Sect. \ref{sec:discussion_velocity_evolution}). We calculated the collision velocities and the sizes of the projectiles that destroy targets of different sizes, and, ultimately, the collisional lifetimes of particles in the disc. By comparing the collisional lifetimes with system ages for each disc, we were able to place constraints on the sizes of the bodies feeding the cascades.

\subsection{Collision velocity}
Debris discs are made up of large numbers of bodies on randomised orbits. We assumed all the discs in our sample to be axisymmetric, such that the orbital angles are uniformly distributed \citep[but see][]{Lovell2026}. Based on simulations of gravitational interactions among planetesimals and based on observations of the Kuiper belt \citep{Ida1992, LissauerStewart1993, Brown2001}, it is reasonable to expect that the distributions of the orbital eccentricity and inclination in the observed debris discs might be close to the Rayleigh distribution \citep[see, however,][]{Zawadzki2026,Chiang2026}. If that is the case, in a disc with an average particle orbital eccentricity, $e$, and inclination, $i$, the average relative velocity with which particles with similar semi-major axes approach each other  \citep{LissauerStewart1993} is given by
\begin{equation}
    v_{\rm rel} = \sqrt{\frac{5}{4} \langle e^2 \rangle + \langle i^2 \rangle} v_{\rm K},
\end{equation}
where $v_{\rm K}$ is the local Keplerian velocity. We further assumed that $\sqrt{\langle e^2 \rangle} \approx 2 \sqrt{\langle i^2 \rangle}$. This is expected if the origin of the inclination dispersion is gravitational stirring by large bodies present in the disc \citep[in the dispersion-dominated regime,][]{Ida1992}. Then, $v_{\rm rel}$ is determined only by the average inclination.

For a Rayleigh distribution of inclinations, the vertical spatial distribution of particles is Gaussian, and the average inclination is related to the disc aspect ratio as $\sqrt{\langle i^2 \rangle} = \sqrt{2} h$ \citep{Matra2019}. Combining all of these equations, we can obtain a simple relationship between $h$ and $v_{\rm rel}$,
\begin{equation} \label{eq:v_rel}
    v_{\rm rel} = 2\sqrt{3} h v_{\rm K}.
\end{equation}

\citet{Zawadzki2026} found that for many of the ARKS discs the vertical structure at mm wavelengths is not, in fact, Gaussian. They reported the aspect ratio corresponding to half-width half-maximum of the vertical structure. We assumed that accounting for a different vertical distribution or the slightly different aspect ratio definition would only change $v_{\rm rel}$ by a factor of an order of unity and, thus, we neglected this factor in favour of Eq. (\ref{eq:v_rel}), which is widely adopted \citep[e.g. by][]{Zawadzki2026}.

With the aspect ratios, stellar masses, and the central disc radii obtained as described in Sect. \ref{sec:data}, $v_{\rm rel}$ ranges between 40\,m\,s$^{-1}$ and 2600\,m\,s$^{-1}$ for the discs in our sample (see Table \ref{table:data}). 
These collision velocities are higher than the escape velocities from the planetesimals feeding the cascades, whose sizes we present further below (the escape velocity is $\sim 10$\,m\,s$^{-1}$ for rocky 10\,km-sized planetesimals). This means that the gravity of the planetesimals does not significantly influence the collision rates and the role of gravitational focusing can be neglected within the collisional cascade.

\subsection{Projectile-to-target mass ratio}
Collisional fragmentation occurs if the impact energy is higher than the critical energy to disrupt the target. Disruptive or catastrophic collisions are those where the largest surviving body after the collision has less than half of the mass of the target. At a fixed collision velocity, $v_{\rm rel}$, there is a minimum mass of a projectile, $m_{\rm c}$, that can collisionally disrupt the target of mass $m_{\rm t}$, and if $m_{\rm c} \ll m_{\rm t}$ \citep[e.g.][]{Krivov2006}, we have
\begin{equation} \label{eq:crit_mass_ratio}
    \frac{m_{\rm c}}{m_{\rm t}} = \frac{2 Q_{\rm D}^*}{v_{\rm rel}^2},
\end{equation}
where $Q_{\rm D}^*$ is the specific critical disruption energy that depends on the target size, composition, and the collision velocity \citep[e.g.][]{Benz1999}. The critical projectile size (radius) is given by $s_{\rm c} = s_{\rm t} ( 2 Q_{\rm D}^* / v_{\rm rel}^2 )^{1/3}$, for the target size, $s_{\rm t}$.

For the specific critical disruption energy, we adopted
\begin{equation} \label{eq:q_d_star}
    Q_{\rm D}^*(s, v_{\rm rel}) = \left( Q_{\rm s} \left( \frac{s}{s_{\rm s}} \right)^{b_{\rm s}} + Q_{\rm g} \left( \frac{s}{s_{\rm g}} \right)^{b_{\rm g}} \right)
    \left( \frac{v_{\rm rel}}{v_{0}} \right)^{b_{\rm v}},
\end{equation}
where $s$ is the particle size (radius), $Q_{\rm s} = Q_{\rm g} = 5\times 10^6$\,erg\,g$^{-1}$ are the reference values in the strength and the gravity regime, respectively, at the reference sizes, $s_{\rm s}=1$\,m and $s_{\rm g}=1$\,km, $b_{\rm s}=-0.37$ and $b_{\rm g}=1.38$ are the slopes of the size-dependence in the strength and the gravity regime, $v_0=3$\,km\,s$^{-1}$ is the reference velocity, and $b_{\rm v}=0.5$ is the slope of the velocity-dependence \citep[all values for basalt;][]{Benz1999, Stewart2009}. For very large bodies ($\gtrsim 100$\,km) the specific energy required to overcome self-gravity exceeds the above $Q_{\rm D}^*$. An additional term can be included to account for this \citep{Leinhardt2009,Lohne2017}. In this work we find planetesimals are small enough to neglect this term.

\begin{figure}
    \centering
    \includegraphics[width=\linewidth]{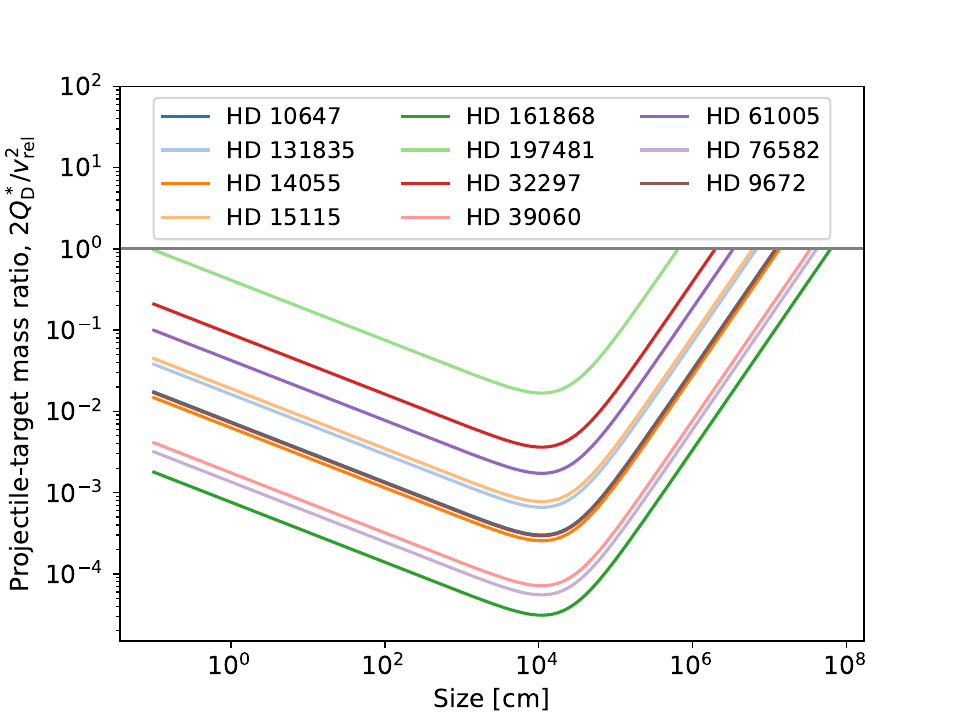}
    \caption{Critical projectile-to-target mass ratio as function of target size for the discs in our sample, obtained using Eqs. (\ref{eq:v_rel}), (\ref{eq:crit_mass_ratio}), and (\ref{eq:q_d_star}), and using data provided in Table \ref{table:data}. The horizontal gray line indicates where $2 Q_{\rm D}^*/v_{\rm rel}^2 = 1$. }
    \label{fig:crit_ratio}
\end{figure}

Figure \ref{fig:crit_ratio} shows the resulting critical projectile-to-target mass ratio as a function of target size for our sample of debris discs. We find that that the critical mass ratios for mm-sized grains are in the range of $10^{-3} - 1$, meaning they indeed fragment in collisions with smaller or at least with similar-size particles. In some discs, for smallest and largest sizes the ratio $2 Q_{\rm D}^* / v_{\rm rel}^2$ is larger than unity. We cut off the curves in Fig. \ref{fig:crit_ratio} for such parameters because Eq. (\ref{eq:crit_mass_ratio}) is no longer valid and the ratio no longer represents the projectile-target mass ratio.

Finally, note that these results depend on the assumed material composition and porosity. For example, if the dust and the planetesimals are icy, they are disrupted more easily \citep{Benz1999} and their critical projectile sizes would be even smaller.

\subsection{Size distribution in steady-state} \label{sec:size_distr}
To model the size distribution in a collisional cascade at the steady-state, we followed the model of \citet{Wyatt2011}, which is similar to the model from \citet{Lohne2008}. The model predicts a size distribution that is a broken power law. A collisional cascade that is at steady-state and in which particle strength depends on particle size as a power law ($Q_{\rm D}^* \propto s^b$) has a size distribution that is also a power law ($n(s) \propto s^{-q}$), with a slope \citep{O'Brien2003} of
\begin{equation} \label{eq:q_b}
    q = \frac{7 + b/3}{2 + b/3}.
\end{equation}
In debris discs, $Q_{\rm D}^*$ scales differently with size for small bodies in the strength regime and for large bodies in the gravity regime \citep[as in Eq. (\ref{eq:q_d_star}),][]{Benz1999,Stewart2009}. Hence, we expect bodies in debris disc cascades to have a broken power-law size distribution, with the break at the transition between the strength and the gravity regime. The transition is at the size of the weakest bodies ($s_{\rm w}$), that is, the size where $Q_{\rm D}^*$ is minimised. For a $Q_{\rm D}^*$ law of the form from Eq. (\ref{eq:q_d_star}) the size of the weakest bodies  \citep{Wyatt2011} is given by
\begin{equation}
    s_{\rm w} = \left( - \frac{Q_{\rm g} b_{\rm g} s_{\rm g}^{-b_{\rm g}}}{Q_{\rm s} b_{\rm s} s_{\rm s}^{-b_{\rm s}}} \right)^{1/(b_{\rm s} - b_{\rm g})},
\end{equation}
where $b_{\rm s} < 0$, making the expression in the parentheses positive. For the dispersal law assumed here, $s_{\rm w}\approx 110$\,m. The differential size distribution is given by
\begin{equation}
    n(s) = \begin{cases}
        K_{\rm s} s^{-q_{\rm s}}, s<s_{\rm w}, \\
        K_{\rm g} s^{-q_{\rm g}}, s>s_{\rm w},
    \end{cases}
\end{equation}
where $q_{\rm s}=3.66$ and $q_{\rm g}=3.03$ are the slopes in the strength and the gravity regime, calculated using Eq. (\ref{eq:q_b}) and the adopted values for $b_{\rm s}$ and $b_{\rm g}$ from Eq. (\ref{eq:q_d_star}), respectively. The normalisation constant $K_{\rm s}$ for the small bodies is calculated by assuming the observed dust mass ($M_{\rm dust}$) is the mass of particles up to $s_{\rm cm}=1$\,cm,
\begin{equation}
    K_{\rm s} = \frac{3 (4-q_{\rm s})}{4 \pi \rho_{\rm s}} s_{\rm cm}^{q_{\rm s}-4} M_{\rm dust}.
\end{equation}
We assume the bulk density of particles is $\rho_{\rm s}=2.6$\,g\,cm$^{-3}$. These assumptions are consistent with how our dust masses are obtained from continuum fluxes \citep[][and see also Sect. \ref{sec:data}]{Marino2026a}.

For large bodies, the normalisation constant $K_{\rm g}$ is calculated from a requirement that at steady-state the mass flux through the cascade is independent of size, and therefore equal in the strength regime and in the gravity regime. Following \citet{Wyatt2011}, but accounting for the velocity dependence of $Q_{\rm D}^*$, we find
\begin{equation}
    K_{\rm g} = K_{\rm s} \sqrt{\frac{1-q_{\rm g}}{1-q_{\rm s}}} \left[ \frac{2}{v_{\rm rel}^2} \left( \frac{v_{\rm rel}}{v_0} \right)^{b_{\rm v}} \right]^{(q_{\rm g}-q_{\rm s})/6} \frac{\left( Q_{\rm s} s_{\rm s}^{-b_{\rm s}} \right)^{(1-q_{\rm s})/6}}{\left( Q_{\rm g} s_{\rm g}^{-b_{\rm g}} \right)^{(1-q_{\rm g})/6}}.
\end{equation}

The size distribution modelled in this way contains a small discontinuity between the two regimes, whose root cause is the fact that targets just above the transition (sizes just above $s_{\rm w}$) are destroyed by projectiles whose size distribution is that of the small bodies in the strength regime \citep{O'Brien2003,Wyatt2011}. Nevertheless, similar analytic broken power-law distributions have been shown to agree well with numerical simulations \citep{O'Brien2003,Lohne2008,Wyatt2011}. In simulations, and we expect in reality, there is no discontinuity at the transition. Instead, the transition between the strength and the gravity regime causes `ripples' or `waves' in the distribution of the large bodies. The amplitude of these waves is similar to the magnitude of the discontinuity in the analytic model. We neglect these waves in this work.

\begin{figure*}
    \centering
    \includegraphics[width=\columnwidth]{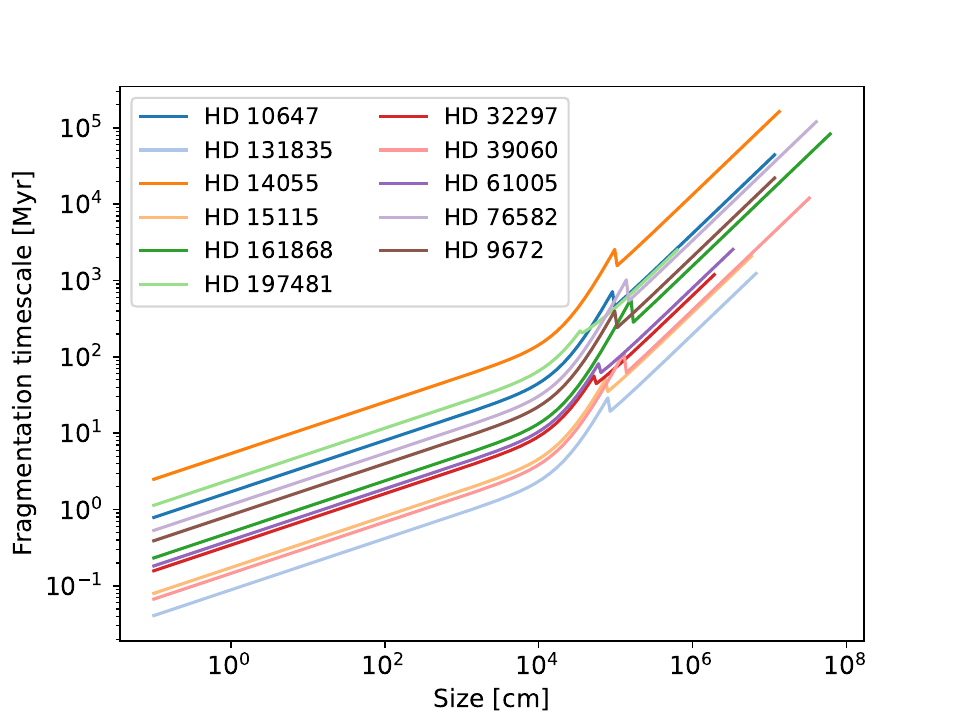}
    \includegraphics[width=\columnwidth]{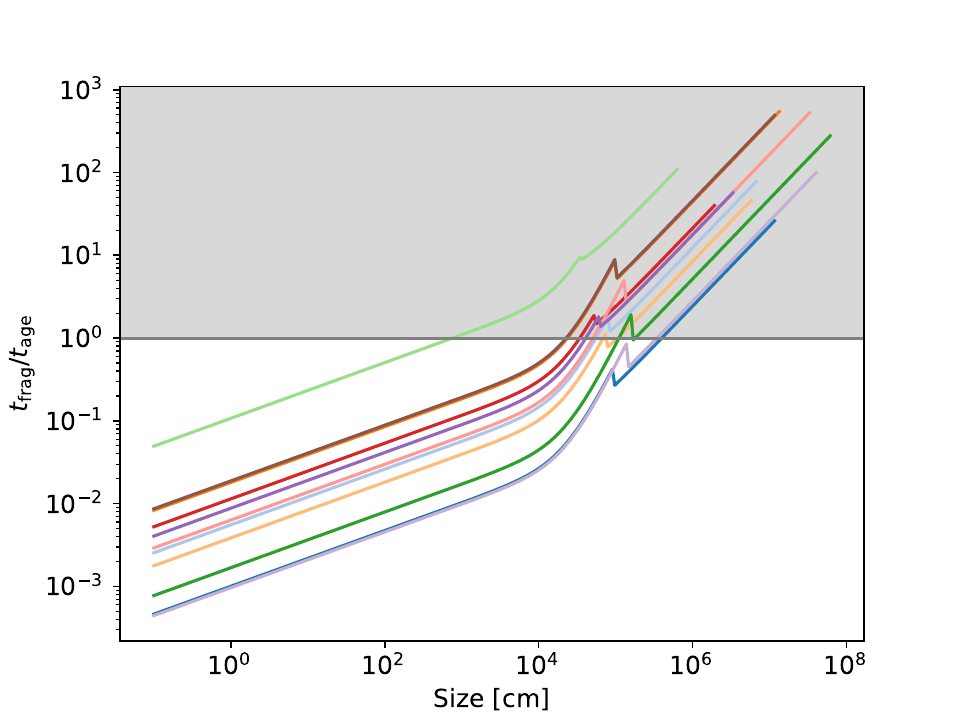}
    \caption{\textit{Left}: Fragmentation timescale as function of particle size. \textit{Right}: Ratio of the fragmentation timescale and the system age for each of the discs. Lines on the right follow the same key legend shown in the panel on the left. Lines are cut off at the upper end at the sizes at which the critical disruption energy becomes too large and the bodies cannot be catastrophically disrupted except in collisions with larger bodies. In the right-hand-side panel, the gray region shows where the fragmentation timescale is longer than system age.}
    \label{fig:t_frag}
\end{figure*}

\subsection{Fragmentation timescale} \label{sec:frag_timescale}
The critical projectile-to-target mass ratios calculated above indicate that the estimated collision velocities are sufficiently high for particles to be fragmented. We calculated the timescale on which bodies of different sizes fragment. For this purpose, we used a simple particle-in-a-box collisional model similar to \citet{Wyatt2007a} and \citet{Wyatt2011}. We assumed all particles are well mixed within the volume of the disc given by
\begin{equation}
    V = 4 \pi r^2 \Delta r h,
\end{equation}
where $r$, $\Delta r$ and $h$ are the radius of the belt centre, its full width and its aspect ratio, respectively, estimated from observations as described in Sect. \ref{sec:data}. Likewise, we assume all particles collide, on average, with the same relative velocity ($v_{\rm rel}$, estimated as described above).

Targets can be destroyed by all projectiles larger than the critical projectile size. The fragmentation rate of a target of size $s_{\rm t}$ is given by
\begin{align}
    R_{\rm frag} = \frac{\pi v_{\rm rel}}{V}  \int_{s_{\rm c}}^{s_{\rm max}} n(s_{\rm p}) (s_{\rm t} + s_{\rm p})^2 \textrm{d}s_{\rm p},
\end{align}
where $s_{\rm max}$ is the radius of the biggest body in the disc. We further assume that $q>3$ so that small sizes dominate the collisional cross-section (this is true for our steady-state distribution). We also use the fact that the critical projectile sizes we found tend to be much smaller than the target sizes ($s_{\rm c} \ll s_{\rm t}$). Under these assumptions, we have
\begin{align} \label{eq:r_frag}
    R_{\rm frag} \approx \frac{\pi v_{\rm rel} }{V} \frac{K_{\rm c}s_{\rm c}^{1-q_c} s_{\rm t}^2}{(q_c-1)},
\end{align}
where the size distribution constant, $K_{\rm c}$, and slope, $q_{\rm c}$, depend on the size distribution around $s_{\rm c}$.\footnote{For a complete solution to the integral without the above assumptions, we refer the reader to \citet{Lohne2008}.} The fragmentation timescale is simply the inverse of the rate,
\begin{equation} \label{eq:t_frag}
    t_{\rm frag} = \frac{(q_c-1) V}{\pi K_{\rm c} v_{\rm rel}} s_{\rm c}^{q_c-1} s_{\rm t}^{-2}.
\end{equation}
We note that in this simple model, the dependence of the fragmentation timescale on the disc scale-height is purely through the critical projectile size, since other occurrences of $h$ get cancelled out.

The results are shown in Fig. \ref{fig:t_frag}. We find that the collisional lifetimes of the observed mm-sized grains are several orders of magnitude shorter than the system ages, as expected (since the lines are below unity at mm sizes in the right-hand-side panel). The dust cannot be primordial (i.e. a remnant from the protoplanetary disc phase) and, instead, it has to have been produced more recently \citep{Backman1993}. The collisional lifetimes are also at least $\sim 100$ times longer than the orbital timescale (see Fig. \ref{fig:t_frag_t_k}).

We also do not show the lifetime (i.e. we cut off the plotted lines) for large particles for which Eq. (\ref{eq:crit_mass_ratio}) evaluates to above unity, as such particles cannot, in fact, be catastrophically disrupted by bodies smaller than themselves. All lines on the right in the right-hand-side panel cross above $t_{\rm frag}/t_{\rm age}=1$, showing that the limiting factor to particles participating in the cascade is indeed their collisional timescale as opposed to them being too strong to be disrupted.

\subsection{Minimum initial disc mass}
A disc in a steady-state collisional cascade loses mass over time. Assuming that mass is the only disc property that changes (e.g. the size distribution slope is fixed), disc mass decreases as \citep{Wyatt2007a}
\begin{equation} \label{eq:m_cascade}
    M_{\rm disc} = M_0 / ( 1 + t/t_0 ),
\end{equation}
where $M_0$ is the initial disc mass and $t_0$ is the initial fragmentation timescale of the largest bodies. The initial disc mass is unknown, but its minimum value can be constrained \citep{Krivov2021,Pearce2025} via
\begin{equation} \label{eq:M_0}
    M_0 \geq 4 |\dot{M}| t_{\rm age},
\end{equation}
where $\dot{M}$ is the current mass loss rate through the cascade,
\begin{equation}
    \dot{M} = - \frac{M(s_{\rm dust})}{t_{\rm frag}(s_{\rm dust})},
\end{equation}
where $M(s_{\rm dust})$ is the mass in particles of size around $s_{\rm dust}$. Here and throughout this work, we estimate the mass in particles of size around some size $s$ as the mass in the unit log-bin of size around $s$,
\begin{equation} \label{eq:m_in_s}
    M(s) = \frac{4\pi\rho_s}{3} K s^{4-q},
\end{equation}
where $K$ and $q$ are the normalisation and the slope of the size distribution around $s$. Using the ARKS dataset, the steady-state size distributions and the fragmentation timescales calculated as outlined in previous sections, adopting $s_{\rm dust}=1$\,mm, we find minimum initial disc masses in the range of 0.1 -- 200\,M$_\oplus$. This is only a rough, order-of-magnitude estimate, as the chosen width of the size bin in $M(s)$ is arbitrary. Had we chosen to define $M(s)$ as the mass from \textmu m-sized to mm-sized grains, the inferred masses would be larger by a factor of $\sim 7$. The choice of $s_{\rm dust}$ is also arbitrary, but it is irrelevant~-- by definition our steady-state collisional cascade model has a mass flow rate that is constant with size. An alternative is to consider small, blowout-sized particles, which are assumed to dominate the discs' fractional luminosities ($f$) with a mass-loss rate  \citep{Matra2017} given by
\begin{align} \label{eq:mdot_blowout}
    \dot{M} &= - \frac{M_{\rm blowout}}{t_{\rm frag}(s_{\rm blowout})} \\
    &= - 1.2 \times 10^3 \frac{\textrm{M}_\oplus}{\rm Myr}
    \left(\frac{r}{\rm au}\right)^{3/2}
    \left(\frac{\Delta r}{\rm au}\right)^{-1}
    f^2
    \frac{L_*}{\textrm{L}_\odot}
    \left(\frac{M_*}{\textrm{M}_\odot}\right)^{-1/2}.
\end{align}
The expression follows from the same  collisional model employed in this work \citep[][]{Wyatt2007a}. \citet{Matra2017} assume a size distribution slope $q=3.5$, different from our $q_{\rm s}=3.66$, but this would change the mass-loss rate only by about a percent (see their Eq. 19). For the blowout-sized particles, there is no dependence on the collision velocity (or on the disc's aspect ratio) because the critical projectile-target size ratio is irrelevant~-- smaller particles are blown out by radiation pressure and blowout-sized particles predominantly collide with equal-sized particles.

\begin{figure}
    \centering
    \includegraphics[width=\columnwidth]{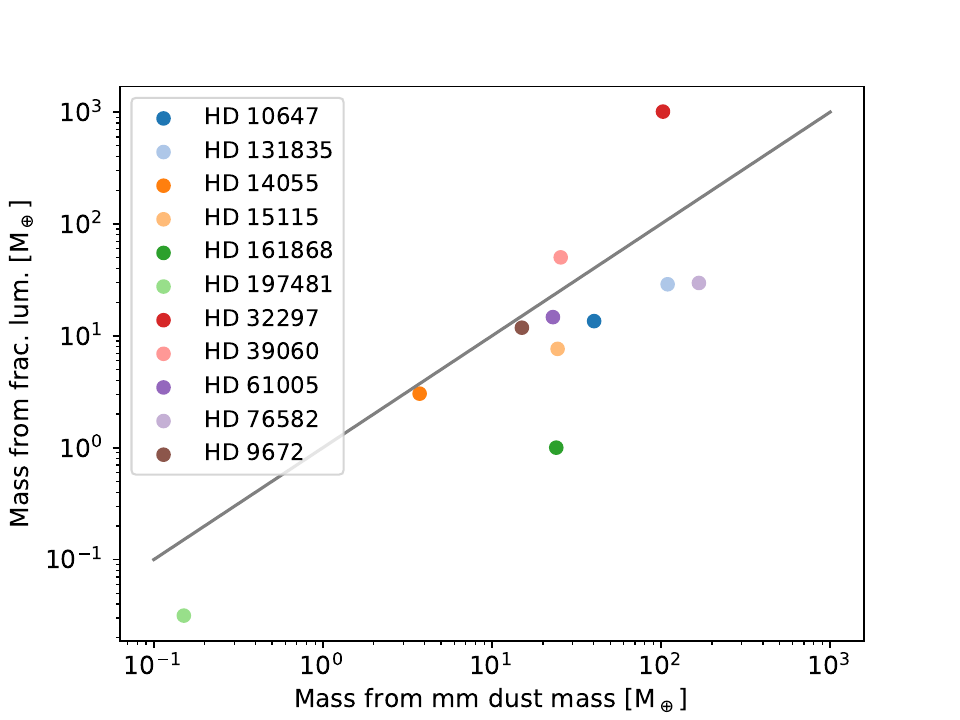}
    \caption{Minimum initial mass in collisional cascade calculated using millimetre dust mass compared to mass calculated using fractional luminosity. Gray line indicates where the two are equal.}
    \label{fig:m_pl_min_initial}
\end{figure}

Minimum initial disc masses estimated from fractional luminosities tend to be somewhat lower than the masses estimated from millimetre dust masses, as shown in Fig. \ref{fig:m_pl_min_initial}. Nevertheless, the two estimates largely agree within an order of magnitude. This is striking in view of the vast simplifications made to relate the measured fractional luminosities with the number and the location of blowout-sized grains and any contribution from sub-blowout grains, and to estimate the mass flow rate at millimetre sizes. The agreement between the two estimates is good even for HD~197481 (AU~Mic) and HD~61005, two low-luminosity stars for which the minimum dust grain size may very likely not be the blowout limit set by radiation pressure, and this points to the need for caution when interpreting these agreements.

\subsection{Initial disc mass and planetesimal size} \label{sec:m_pl}
Further constraints on the bodies hiding in the disc can be achieved, albeit with some additional assumptions. In this section we extrapolate the size distribution of larger bodies from the observed dust masses assuming steady-state collisional cascades (see Sect. \ref{sec:size_distr}) and we estimate the sizes at the upper end of the cascades, that is, sizes of the planetesimals feeding the cascades.

First, we note that the minimum initial disc mass from Eq. (\ref{eq:M_0}) corresponds to the initial fragmentation timescale of the largest bodies being equal to the system age \citep[this follows simply from minimisation of the initial disc mass over the initial fragmentation timescale;][]{Pearce2025}. Indeed, there are strong arguments to assume that the planetesimals at the top of a steady-state cascade, of radius $s_{\rm pl}$, are those whose fragmentation timescale equals the system age, and they represent the lower limit on the size of primordial bodies in the disc \citep{Wyatt2002}. Primordial planetesimals in an observed disc cannot be much smaller because smaller planetesimals fragment too quickly and any initial disc mass would have been depleted by the estimated age of the systems\footnote{This can also be verified by expressing the initial disc mass required to explain the current dust mass with largest planetesimals of some size, $s<s_{\rm pl}$, in terms of the initial disc mass required for $s=s_{\rm pl}$, using Eqs. (\ref{eq:m_cascade}), (\ref{eq:m_in_s}) and (\ref{eq:t_frag}). Smaller planetesimals require larger initial disc masses to explain dust detections. As first noted by \citet{Wyatt2002}, for $s$ that is only moderately lower than $s_{\rm pl}$ (e.g. for $s=s_{\rm pl}/4$ in the idealised cascade with $Q_{\rm D}^*=$\,const and $q=3.5$), the required initial disc mass becomes infinite.}. On the other hand, larger planetesimals may exist in each disc, but in the steady-state model their collisions are happening too infrequently to have produced a lot of dust, and they still have their primordial distribution \citep{Lohne2008}. Such larger bodies may contribute to the cascade through non-disruptive, erosive collisions, and this possibility is explored in a companion paper by \citet{Chiang2026}.

Second, we recall that the relevant fragmentation timescale is that at the start of the cascade, that is, at time $t=0$. As already noted above, the fragmentation timescale at the start is a function of the initial disc mass. Additionally, the initial size distribution may differ from the one at steady-state \citep{Lohne2008,Schlichting2013,Krivov2018a,Krivov2021}. Here, we follow \citet{Lohne2008} and \citet{Krivov2021} and assume that the size distribution of primordial planetesimals was a power law characterised by $q_{\rm p}$ and $K_{\rm p}$. The initial disc mass obtained from this analysis will therefore differ from the one obtained in the previous section, where the underlying assumption of Eq. (\ref{eq:m_cascade}) is that the size distribution is fixed to the steady-state distribution from the start.

We find the size of planetesimals feeding a cascade by requiring that $t_{\rm frag}(s_{\rm pl}) = t_{\rm age}$ and that the mass in these planetesimals, $M(s_{\rm pl})$, equals the current, steady-state mass in these bodies \citep[see also Sect. 4.3 of][]{Pearce2025}.
For each disc, we calculate the present-day, steady-state size distribution, as described in Sect. \ref{sec:size_distr}. For an arbitrary planetesimal size, $s$, we can then calculate the present-day mass in a unit log-bin of size using Eq. (\ref{eq:m_in_s}). We assume that this mass has not changed significantly since $t=0$, and for a given $q_{\rm p}$ we can calculate the primordial size distribution normalisation constant $K_{\rm p}$. Then, we can calculate the fragmentation timescale of these particles at $t=0$. The calculation of this fragmentation timescale is analogous to the calculation of the present-day fragmentation timescale presented in Sect. \ref{sec:frag_timescale}, except that here we assume $K_{\rm c}=K_{\rm p}$ and $q_{\rm c}=q_{\rm p}$ in Eq. (\ref{eq:t_frag}). We assume the other disc properties were otherwise the same in the past, that is, the spatial distribution of the planetesimals has remained unchanged over the stellar lifetime. Finally, we iteratively solve for the size, $s=s_{\rm pl}$, for which $t_{\rm frag}(s) = t_{\rm age}$.

\begin{figure}
    \centering
    \includegraphics[width=\columnwidth]{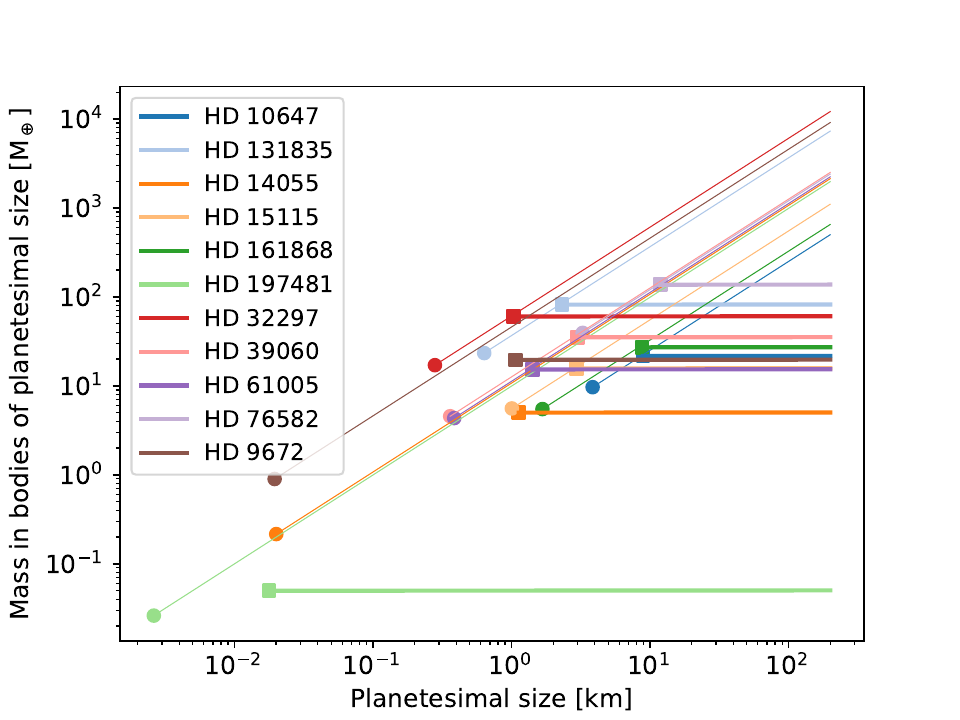}
    \caption{Planetesimal sizes feeding the cascades assuming primordial power-law size distribution with slope $q_{\rm p}=3$ (circles) and its extrapolation up to 200\,km (thin lines), and likewise for $q_{\rm p}=4$ (squares and thick lines), plotted against the corresponding mass inside unit log bin of size around the planetesimal size. For a size distribution with slope $q_{\rm p}=4$ there is equal mass in each log bin of size.}
    \label{fig:m_pl}
\end{figure}

The resulting planetesimal sizes and the masses contained in these planetesimals are shown in Fig. \ref{fig:m_pl} for $q_{\rm p}=3$ (circles) and $q_{\rm p}=4$ (squares). Overall, the planetesimal sizes range from a few metres to ten kilometres, and the total masses of these planetesimals are less than 200\,M$_\oplus$. If we assumed the primordial size distribution was the same as the present-day steady-state size distribution, the inferred planetesimal sizes would be those for which $t_{\rm frag}(s) = t_{\rm age}$ in the right-hand panel of Fig. \ref{fig:t_frag}, and their values would fall between the two estimates shown in Fig. \ref{fig:m_pl} (because $3 < q_{\rm s}, q_{\rm g} < 4$).

From these $s_{\rm pl}$ and $q_{\rm p}$ values for each disc, we were also able to extrapolate the hypothetical primordial size distributions up to 200\,km, shown as the lines in Fig. \ref{fig:m_pl}. These large planetesimal sizes are predicted by planetesimal formation models \citep{Johansen2015,Simon2016,Schafer2017}, but their existence is not required to explain the observations \citep[because their $t_{\rm frag}(s) > t_{\rm age}$;][]{Krivov2021}.

In contrast, our assumed primordial size distributions require some planetesimals smaller than $s_{\rm pl}$ to be initially present. The primordial power-law distributions are assumed to extend down to at least the critical projectile size for $s_{\rm pl}$, and the critical projectile sizes are about 2 to 20 times smaller than $s_{\rm pl}$. If such smaller sizes do not exist at $t=0$, or their number is smaller than implied by these power laws, the initial fragmentation timescales would be longer and the planetesimals feeding the cascade would need to be smaller than values of $s_{\rm pl}$ shown in Fig. \ref{fig:m_pl}. For example, consider the extreme case of an initial size distribution that is a Dirac delta function centred at the parent planetesimal size. We can solve for this planetesimal size in a manner similar to above, and this yields particle radii of only 10\,cm -- 10\,m.

Putting this caveat aside, we explored the relationship between our inferred planetesimal sizes shown in Fig. \ref{fig:m_pl} and stellar masses and disc radii, the two quantities we have in our dataset, which could influence the initial planetesimal mass function \citep[e.g.][]{Gerbig2023}. We did not find any correlations.

\section{Constraints on stirrers} \label{sec:stirrers}
In this section, we present the constraints we set on the sizes and surface densities of large bodies whose gravitational interactions with planetesimals and dust could result in the observed disc scale-heights. We assume that: (1) stirrers are embedded within the disc; (2) stirrers fully formed early on, during the protoplanetary disc stage or quickly after; (3) planetesimals and dust have been stirred starting from a negligibly low excitation level, for the duration of the system's lifetime; and (4) excitation of stirrers is at all times lower than that of planetesimals and dust. We discuss these assumptions and alternative theories in Sects. \ref{sec:discussion_velocity_evolution} and \ref{sec:alternative}.

If a disc starts with a very low excitation, the gravitational interactions with embedded stirrer(s) initially occur in a shear-dominated (or sub-Hill) regime where relative velocities are primarily determined by Keplerian shear. This is up until particle eccentricities increase sufficiently to drive higher relative velocities, and the disc enters the dispersion-dominated (or super-Hill) regime. Debris discs are typically assumed to be in the dispersion-dominated regime \citep[e.g.][]{Quillen2007,Pan2012,Matra2019}, and so we first present results making the same assumption. We then verify that none of the discs in our sample are currently in the shear-dominated regime if collisional damping is neglected, and finally we consider the role of collisional damping.

\begin{figure*}
    \centering
    \includegraphics[width=\columnwidth]{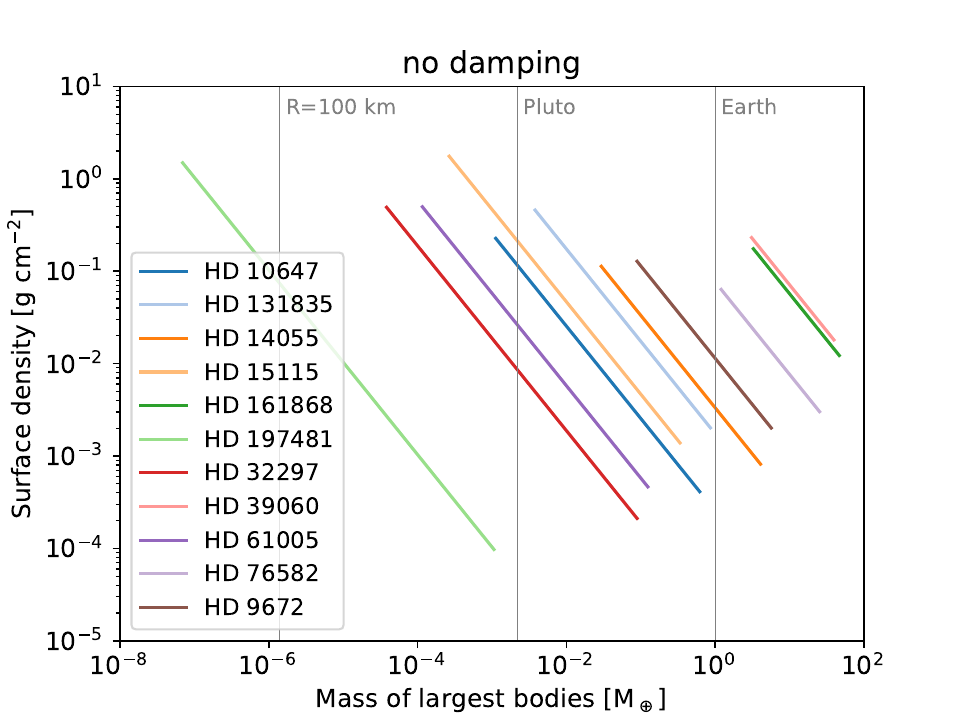}
    \includegraphics[width=\columnwidth]{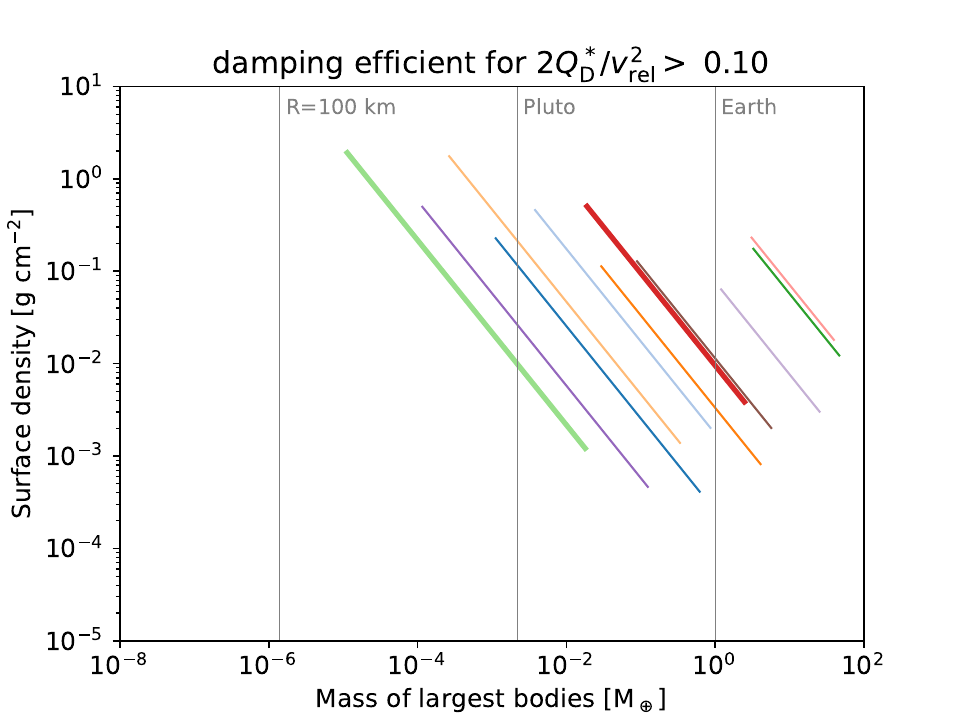}
    \caption{Constraints on the surface density of stirrers as function of their mass, assuming stirring in dispersion-dominated regime. Left: No collisional damping. The product of the surface density and the mass is constrained by the stirring timescale and given by Eq. (\ref{eq:m_sigma}). The lines indicate the corresponding possible combinations of stirrers' surface density and mass. The endpoints of lines are determined by the minimum and the maximum surface density, and the minimum mass, given by Eqs. (\ref{eq:sigma_min}), (\ref{eq:sigma_max}) and (\ref{eq:m_min}), respectively. Right: Stirring is balanced by collisional damping if damping is faster than fragmentation, in which case the constraints are obtained by equating the stirring rate with the damping rate from Eq. (\ref{eq:r_damp}). Damping is efficient in only two discs (HD~197481 and HD~32297), shown by the thick lines.}
    \label{fig:m_sigma_disp}
\end{figure*}

\subsection{Gravitational stirring in dispersion-dominated regime} \label{sec:stir_disp}
To constrain properties of the stirrers in the dispersion-dominated regime, we largely follow the approach of \citet{Matra2019}. As in their work, we use the semi-analytic stirring rates from Eqs. (4.1) and (4.2) from \citet{Ida1993} to determine the time evolution of particle eccentricities and inclinations. For details, see Appendix \ref{sec:derivations}. Given the system age and present-day inclination dispersion, the product of stirrer mass, $M_{\rm stir}$, and surface mass density, $\Sigma_{\rm stir}$, of stirrers can be expressed as
\begin{equation} \label{eq:m_sigma}
    M_{\rm stir} \Sigma_{\rm stir} = 0.25 G^{-1/2} M_*^{3/2} r^{-1/2} t_{\rm age}^{-1} i_{\rm rms}^4,
\end{equation}
where $G$ is the gravitational constant and $i_{\rm rms} = \sqrt{\langle i^2 \rangle} = \sqrt{2} h$ is the root mean square of the particle inclinations. This formula is approximately valid in the case of a large number of small stirrers, as well as for a single stirrer with an effective surface density determined by the stirrer's mass and the width of the zone in which its stirring is effective \citep[stirring width;][]{Ida1993}, up to a factor of an order of unity, which depends on whether the bodies are stirring themselves or whether they are stirred by a larger body or bodies. 

There needs to be at least one whole stirrer within the disc and within its own stirring width, and so there is a minimum (effective) surface density,
\begin{equation} \label{eq:sigma_min}
    \Sigma_{\rm stir} > \frac{M_{\rm stir}}{2 \pi r \textrm{\,min}(\Delta r, \Delta r_{\rm stir})},
\end{equation}
 
where $\Delta r_{\rm stir} \approx 4 r h_{\rm Hill} \sqrt{12+20/3 (i_{\rm rms}/h_{\rm Hill})^2}$ is the stirring width \citep{Ida1993} and 
\begin{equation} \label{eq:h_hill}
    h_{\rm Hill} = \left( \frac{M_{\rm stir}}{3 M_*} \right)^{1/3}
\end{equation}
is the reduced Hill radius. 

Furthermore, for a fixed size of the stirrer(s), there is a maximum excitation that can be achieved, corresponding to $v_{\rm rel} \sim v_{\rm esc} = \sqrt{2G M_{\rm stir}/R_{\rm stir} }$, the escape velocity of the stirrer(s) \citep{Safronov1972,Goldreich2004,Schlichting2014}, where $R_{\rm stir}$ is the stirrer radius. For higher relative velocities gravitational focusing becomes inefficient. Particles are then more likely to physically collide with the stirrer than they are to gravitationally scatter. This yields a constraint on the minimum mass of the largest bodies in the disc \citep[e.g.][]{Matra2019,Marino2021,Terrill2023}, expressed as
\begin{equation} \label{eq:m_min}
    M_{\rm stir} > 2.5 \rho_{\rm stir}^{-1/2} M_*^{3/2} r^{-3/2} i_{\rm rms}^3.
\end{equation}
For the stirrers, we assumed $\rho_{\rm stir}=2$\,g\,cm$^{-3}$, noting that we find stirrers' sizes range from minor bodies to Neptune-sized planets, so their bulk density might be diverse in reality. We added another constraint: the total mass of stirrers in the belt implied by $\Sigma_{\rm stir}$ cannot be larger than the solid mass contained inside the systems' protoplanetary discs, $M_{\rm PPD}$. This gives
\begin{equation} \label{eq:sigma_max}
    2 \pi r \Delta r \Sigma_{\rm stir} < M_{\rm PPD}.
\end{equation}
Here, we assume the spatial distribution of stirrers has not evolved over the system's lifetime and it is similar to the spatial distribution of dust. We can estimate $M_{\rm PPD}$ using the standard dust-to-gas ratio of 0.01 and a total PPD mass of 0.1 of the stellar mass \citep{Williams2011}. We used the solid mass because the stirrers we found are mostly smaller than 20\,M$_\oplus$, corresponding to sub-Neptunes, which are expected to have less than a few percent of their mass in their atmospheres.

Together, the four constraints in Eqs. (\ref{eq:m_sigma}), (\ref{eq:sigma_min}), (\ref{eq:m_min}), and (\ref{eq:sigma_max}) yield finite ranges of $M_{\rm stir}$ and $\Sigma_{\rm stir}$ shown in the left panel of Fig. \ref{fig:m_sigma_disp}. There is a wide dispersion in the mass of stirrers inferred, from planetesimal-sized bodies for HD\,197481 (but see Sect. \ref{sec:coll_damp}) to several tens of Earth mass for HD\,39060 and HD\,161868. Overall, these masses are realistic, such a wide range of large bodies may indeed exist in different planetary systems, albeit this raises the question of what planet formation processes create the dispersion. Furthermore, we find that Eq. (\ref{eq:m_min}) should be used with caution. For all discs except for HD\,197481 we find that the condition on the maximum disc mass from Eq. (\ref{eq:sigma_max}) is more stringent than the lower limit on stirrers' individual masses from Eq. (\ref{eq:m_min}). This means that for almost all discs in the sample the smallest possible stirrers from Eq. (\ref{eq:m_min}) need to be unrealistically numerous, with their total mass larger than the mass available in the solids in their nascent protoplanetary disc to stir the discs within system age. 

\begin{figure}
    \centering
    \includegraphics[width=\columnwidth]{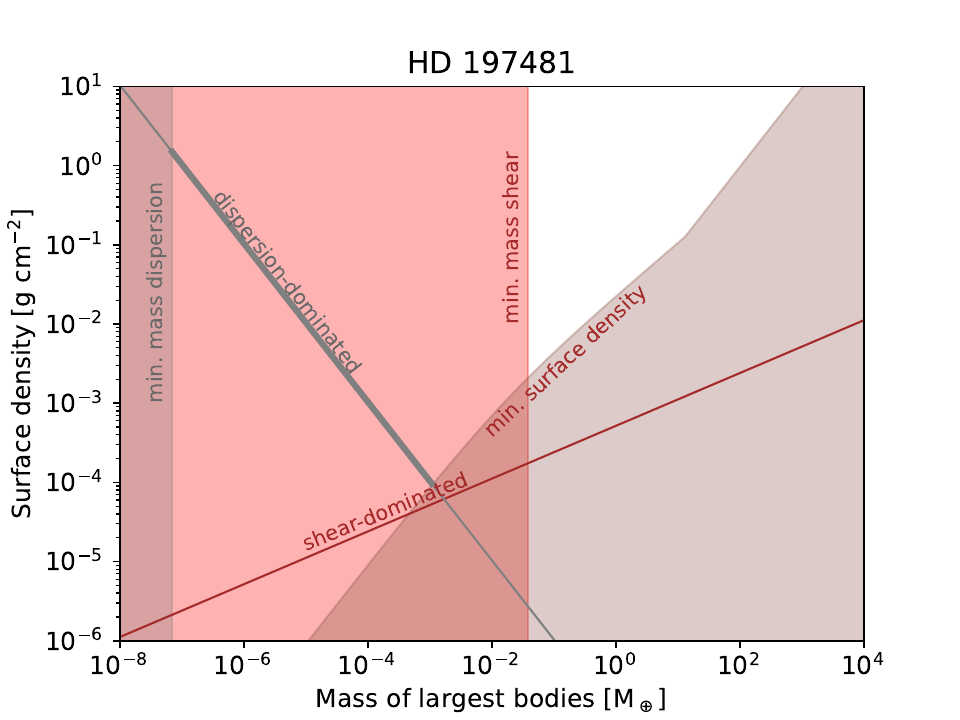}
    \caption{Constraints on surface density and mass of stirrers assuming stirring in the dispersion-dominated regime compared to the shear-dominated regime for HD~197481, ignoring collisional damping. The gray line for the dispersion-dominated regime is determined by Eq. (\ref{eq:m_sigma}) and the red line for the shear-dominated regime by Eq. (\ref{eq:m_sigma_shear}). The minimum surface density condition given by Eq. (\ref{eq:sigma_min}) applies to both regimes. In the shear-dominated regime, the minimum mass is determined by the condition that the stirring is indeed in this regime, given by Eq. (\ref{eq:m_min_shear}). For this disc, when damping is neglected, the left endpoint for the dispersion-dominated regime is determined by the minimum mass from Eq. (\ref{eq:m_min}), but note the condition on the maximum surface density from Eq. (\ref{eq:sigma_max}) is more stringent for most discs.}
    \label{fig:m_sigma_shear}
\end{figure}

\subsection{Gravitational stirring in shear-dominated regime} \label{sec:stir_shear}
A similar semi-analytic model describes stirring in the shear-dominated regime \citep[Eq. 4.7 in][]{Ida1993}, with a corresponding constraint,
\begin{equation} \label{eq:m_sigma_shear}
    M_{\rm stir}^{-1/3} \Sigma_{\rm stir} \approx G^{-1/2} M_*^{1/6} t_{\rm age}^{-1} r^{-1/2} \textrm{ln} \frac{i_{\rm rms}}{i_0},
\end{equation}
where $i_0$ is the initial inclination of particles. For details, see Appendix \ref{sec:derivations}.
We assume that $i_0=10^{-4}$ and note that the dependence of our results on it is only logarithmic. Reducing $i_0$ by several orders of magnitude only makes $M_{\rm stir}^{-1/3} \Sigma_{\rm stir}$ increase by a factor of a few.

The minimum surface density given by Eq. (\ref{eq:sigma_min}) that applies here is same as in the dispersion-limited regime. However, the minimum mass of the stirrers is different. For the disc to be in the shear-dominated regime, $i_{\rm rms} \lesssim h_{\rm Hill}$. From Eq. (\ref{eq:h_hill}) it follows that
\begin{equation} \label{eq:m_min_shear}
    M_{\rm stir} > 3 M_* i_{\rm rms}^3.
\end{equation}
This is equivalent to the requirement that the velocity dispersion is not larger than the Hill velocity, whereas in the dispersion-dominated regime the requirement is that it cannot be larger than the escape velocity of the stirrers.

We find that there are no combinations of $\Sigma_{\rm stir}$ and $M_{\rm stir}$ for which all three of the conditions given by Eqs. (\ref{eq:sigma_min}), (\ref{eq:m_sigma_shear}), and (\ref{eq:m_min_shear}) are satisfied, for any of the discs in our sample. We illustrate this by showing the constraints for the disc with the lowest aspect ratio in Fig. \ref{fig:m_sigma_shear}. The red line shows the constraint from Eq. (\ref{eq:m_sigma_shear}). Along this line, either the mass of individual stirrers is too low to be consistent with the shear-dominated regime at the given excitation, or the surface density is too low for the stirring width to contain one whole stirrer. 
We conclude that none of our discs are in the shear-dominated regime.

We can also test if neglecting the brief shear-dominated phase affects our results in the dispersion-dominated regime. We can calculate $t_{\rm Hill}$, the time it takes to stir $i_{\rm rms}$ to $h_{\rm Hill}$. The constraint $\Sigma M$ calculated in the previous section should change by a factor $\sim (t_{\rm age}-t_{\rm Hill})^{-1} (i_{\rm rms}^4 - h_{\rm Hill}^4) / (t_{\rm age}^{-1} i_{\rm rms}^4)$. We find that $t_{\rm Hill}/t_{\rm age}$ is less than a percent and $i_{\rm rms}/h_{\rm Hill}$ at least a few, making this factor very close to unity for all discs in our sample.

A caveat to these findings is that gravitational scattering at low relative velocities is more complex than in the dispersion-regime \citep[see the discussion in Sect. 4.3 of][]{Goldreich2004}. The semi-analytic approach is less accurate in the shear-dominated regime than in the dispersion-dominated regime, when compared with numerical simulations \citep{Ida1992, Stewart2000}.

\subsection{Collisional damping} \label{sec:coll_damp}
Gravitational stirring might be balanced by collisional damping \citep{Brahic1977, Hornung1985, Pan2012}, a process in which the loss of kinetic energy in inelastic collisions tends to reduce the velocity dispersion (and thus the eccentricity and the inclination dispersion) of dust and/or planetesimals. The damping rate for hard, indestructible targets of size $s_{\rm t}$ can be estimated \citep{Hornung1985,Jankovic2024} as
\begin{align} \label{eq:r_damp}
    R_{\rm damp} &= \frac{\pi v_{\rm rel}}{V}  \int_{s_{\rm min}}^{s_{\rm max}} n(s_{\rm p}) (s_{\rm t} + s_{\rm p})^2 \frac{2 s_{\rm t}^3 s_{\rm p}^3}{(s_{\rm t}^3 + s_{\rm p}^3)^2} \textrm{d}s_{\rm p} \\
    &\approx \frac{\pi v_{\rm rel}}{V} 2 K_{\rm t} s_{\rm t}^{3-q_{\rm t}},
\end{align}
where $K_{\rm t}$ and $q_{\rm t}$ are the normalisation and the exponent of the size distribution near $s_{\rm t}$. The approximation holds for power laws with an exponent of $q<4$, for which large bodies contain most of the mass. The expression is equivalent to Eq. 12 from \citet{Pan2012}.

To revise our constraints on $\Sigma_{\rm stir} M_{\rm stir}$ of stirrers in the dispersion-dominated regime, we can derive an equation analogous to Eq. (\ref{eq:m_sigma}) by equating the stirring rate to the collisional damping rate \citep{Pan2012}, for a dust size that contributes most to the observed flux at millimetre wavelengths, $s_{\rm t}=1$\,mm. However, damping has to act within the collisional lifetime of dust particles. Dust particles are part of the collisional cascade, continually produced in collisions of larger particles and continually destroyed in collisions with smaller ones. Damping is most efficient in collisions with equal-sized particles. If particles collide with and get disrupted by smaller particles much more frequently than they collide with equal-sized particles, collisional lifetime is shorter than the damping timescale, and damping is overall inefficient \citep{Jankovic2024}.

Therefore, to revise our constraints on $\Sigma_{\rm stir} M_{\rm stir}$ of stirrers, we require that $t_{\rm damp} < t_{\rm frag}$, where $t_{\rm damp}=1/R_{\rm damp}$. Assuming $K_{\rm c}=K_{\rm t}$ and $q_{\rm c}=q_{\rm t}=q$, the ratio of the timescales is given by 
\begin{equation}
    \frac{t_{\rm frag}}{t_{\rm damp}} = 2 (q-1) \left(\frac{2 Q_{\rm D}^*}{v_{\rm rel}^2}\right)^{(q-1)/3}.
\end{equation}
In the strength regime relevant for the dust ($q=q_{\rm s}=3.66$), the damping timescale is shorter than the fragmentation timescale if the critical projectile-to-target mass ratio $2 Q_{\rm D}^*/v_{\rm rel}^2 \gtrsim 0.1$. We verify this by also numerically integrating Eqs. (\ref{eq:r_damp}) and (\ref{eq:r_frag}). Only two of our discs fulfil this condition (HD~197481 and HD~32297). For these two discs we replace the constraints on $\Sigma_{\rm stir} M_{\rm stir}$ from Eq. (\ref{eq:m_sigma}) with an analogous expression derived from equating the stirring and the damping rate. We show the results in the right-hand-side panel of Fig. \ref{fig:m_sigma_disp}, in which these two discs are shown with thick lines.

In these calculations, we assumed that collisional damping acts on the observable dust particles. The critical specific energy for fragmentation, $Q_{\rm D}^*$, is a non-monotonic function of size (see Fig. \ref{fig:crit_ratio}), and so it may occur that collisions among dust grains are not damping, but that collisions among their parent bodies are. For the discs in our sample, this is not the case. Assuming the parent bodies are those bodies whose fragmentation timescale is equal to the system age (i.e. for the planetesimals feeding the cascades, whose properties we constrained in Sect. \ref{sec:parent_bodies}), we numerically calculated their collisional damping timescales for $q_{\rm p}=3$ and $q_{\rm p}=4$ and found them to be longer than the system age for each disc.

Fragmenting collisions among the parent bodies may contribute to their velocity evolution and thus influence the stirrer properties constrained here. Models that consider the full velocity distribution, rather than only a characteristic collision velocity, show that fragmentation depends non-trivially on the orbital element distributions of both targets and projectiles, potentially affecting the velocity dispersion \citep{Jankovic2024}. However, these models predict only slow evolution of the average particle inclination, and further work is required to distil their results into prescriptions suitable for constraining disc parameters in simple models such as ours.

Dynamical friction also cannot damp parent bodies' velocities. The inferred collision velocities are higher than the escape velocity of bodies in the cascade for each of the discs. Therefore, the rate of cooling by dynamical friction is equal to the rate of collisional damping \citep{Goldreich2004,Pan2012}, and the same conclusions apply as above.

We also checked how collisional damping changes our conclusions so far for the shear-dominated regime. We derive an equation analogous to Eq. (\ref{eq:m_sigma_shear}) by equating the stirring rate from \citet{Ida1993} with the collisional damping rate, and we apply the new constraints on $\Sigma_{\rm stir}$ and $M_{\rm stir}$ for the two discs in which collisional damping may be acting. We find that only HD~32297 could be stirred in the shear-dominated regime (the red line in Fig. \ref{fig:m_sigma_shear} moves up into the allowed white region for this disc). The possible masses of the stirrers range from $\sim 3$ to $10$\,M$_\oplus$. 

In summary, with a simple prescription for the efficiency of collisional damping, we find that velocity damping from mutual collisions of dust grains balances out gravitational stirring in only two of the eleven discs in our sample (HD\,197481 and HD\,32297). Taking this into account results in new constraints on the properties of stirrers for these two discs, as shown in the right-hand-side panel of Fig. \ref{fig:m_sigma_disp}. The total dispersion in the masses of stirrers in the sample is reduced compared to the damping-free calculations in Sect. \ref{sec:stir_disp}, but still significant. The stirring occurs in the dispersion-dominated regime for all of the discs, except, possibly, for HD~32297.

\section{Discussion} \label{sec:discussion}

\subsection{Debris disc inventories and masses}

\subsubsection{Planetesimals} \label{sec:discussion_planetesimals}
Our collisional modelling shows that the debris discs in our sample contain primordial planetesimals with minimum sizes ranging from $\sim 10$\,m to $10$\,km and minimum masses of $\sim 10^{-2}$ to $10^2$\,M$_\oplus$ (see Figs. \ref{fig:m_pl_min_initial} and \ref{fig:m_pl}).
These bodies feed the presumed steady-state collisional cascades producing the observed dust.
Six of our discs (HD~9672, HD~15115, HD~39060, HD~61005, HD~131835, and HD~197481) were previously modelled by \citet{Krivov2021}.
Despite differences in the modelling and adopted disc properties, our minimum disc masses agree well with theirs for four systems.
For HD~9672 and HD~197481 our masses are about an order of magnitude smaller.

HD~197481 (AU Mic) seems to be an outlier in general. Its very low planetesimal and stirrers' mass appears to be driven by its extremely low aspect ratio. 
It should be noted, however, that the minimum disc mass we derive from fractional luminosity is independent of the aspect ratio, and it is even smaller than the one we derive from millimetre dust mass, and also consistent with a previous estimate of collisional mass of $\sim0.01$\,M$_{\oplus}$ from \citet{Strubbe2006}. A stochastic catastrophic disruption event has previously been proposed \citep{Chiang2017} as an explanation for time-variable features observed in this disc at short wavelengths \citep{Boccaletti2015}. This offers an alternative explanation for this system's unusual properties.

As in \citet{Krivov2021}, we find that the planetesimals feeding the cascades have total masses smaller than $\sim$\,$100$\,M$_\oplus$. Extrapolating the primordial size distribution up to 200\,km with a relatively flat size distribution leads to unrealistically large disc masses, over $1000$\,M$_\oplus$, an issue known as the `debris disc mass problem' \citep{Krivov2021}.

Recently, \citet{Marshall2026} found extrapolated disc masses that are smaller than 1000\,M$_\oplus$ for all 11 discs in our sample. The difference from our work seems to come mainly from the dust masses in their work being about 10 times smaller, which can partly be explained by different assumptions about the dust size distribution and composition.

\subsubsection{Stirrers}
Our results show that minimum-mass stirrers defined by Eq. (\ref{eq:m_min}) 
would need to have a total mass larger than the solid mass of their primordial protoplanetary disc for all but one disc.
This stands in contrast to findings of \citet{Zawadzki2026}, who find all but three of the discs could have been stirred by these bodies with reasonable disc masses. The difference comes from 
that
in the definition of the stirring timescale. To calculate the stirring timescale (but not for the rest of their calculations), \citet{Zawadzki2026} employed the model of \citet{Krivov2018b} which only considers if a disc can be stirred sufficiently for bodies to start disrupting, that is, up to a critical fragmentation velocity of 30\,m\,s$^{-1}$. In this work, we consider if bodies are stirred up to an average relative velocity implied by the observed aspect ratios, which we find to be in the range of 40\,m\,s$^{-1}$ to 2600\,m\,s$^{-1}$.

The sizes of the stirrers we infer mostly range from the size of Pluto to sub-Neptunes. This implies that most discs cannot be stirred to their observed aspect ratios by the $200$\,km-sized planetesimals expected from planetesimal formation theory,
although stirring by those planetesimals would be strong enough to trigger the collisional cascade \citep{Krivov2018b}.
However, the stirrers being large does not necessarily mean that the stirrers have been accreting material beyond the protoplanetary disc phase, or that delayed stirring \citep{Kenyon2002b, Kenyon2004a, Kenyon2008, Kenyon2010} has occurred. The stirrers could also have formed early on, inside the protoplanetary disc, consistent with our assumptions. Analyses of the structure of observed protoplanetary discs suggest that sub-Neptunes already exist early on at the orbital distances of 10s of au, comparable to debris disc sizes, albeit from theoretical arguments it appears challenging to prevent their inward migration and keep them at large orbital separations \citep[e.g.][]{Lodato2019}.

\subsubsection{Connection between planetesimals and stirrers}
\begin{figure}
    \centering
    \includegraphics[width=\columnwidth]{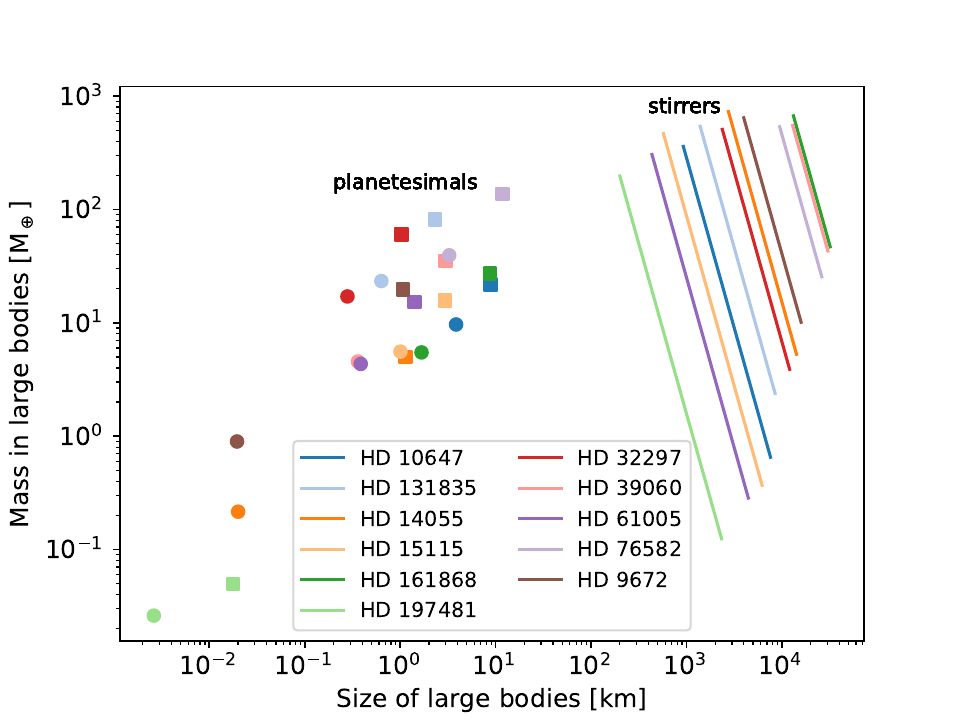}
    \caption{Joint constraints on sizes of planetesimals feeding the collisional cascades (circles for $q_{\rm p}=3$ and squares for $q_{\rm p}=4$, the same as in Fig. \ref{fig:m_pl}) and on the stirrers exciting particles' inclinations (solid lines). The constraints on the stirrers show plausible combinations of their total mass and their individual size, and not a size distribution. These constraints account for collisional damping of mm-sized dust in discs in which it is estimated to be efficient.}
    \label{fig:together}
\end{figure}

In Fig. \ref{fig:together}, we show the full inventories of the debris discs modelled in this work. For the stirrers we convert the surface density constraints into their total mass within each disc, $\Sigma_{\rm stir} 2\pi r \Delta r$. Recall that for the stirrers we constrain the product of their surface density and their mass -- stirrers' total (and individual) masses may lie anywhere along the lines shown in Fig. \ref{fig:together}.

We find that stirrers do not necessarily contain most of the mass within a disc -- for some discs a part of the line (showing possible stirrers' total mass) in Fig. \ref{fig:together} falls below the circles/squares (showing possible planetesimals' total mass). This is analogous to how \citet{Zawadzki2026} find that some of their stirring disc masses are less than the collisional disc masses inferred by \citet{Krivov2021}. Similarly, \citet{Pearce2025} find that the narrow belt of Fomalhaut may contain Pluto-sized stirrers, but if so their total mass could only be only a small fraction of the belt's mass, or otherwise gravitational scattering would make the belt radially broader than it is. Such combinations of stirring and collisional constraints are fully valid: for bodies to dominate stirring it is the product $\Sigma_{\rm stir} M_{\rm stir}$ that needs to be larger, rather than just the total surface density or total mass \citep{Ida1993,Quillen2007,Krivov2018b}.

Throughout this work, we assume that the planetesimals and the stirrers are two distinct populations, with planetesimals feeding the collisional cascades and stirrers providing dynamical excitation. Figure \ref{fig:together} is consistent with this assumption, and indeed the relatively small planetesimals cannot produce the observed disc aspect ratios. However, a companion paper by \citet{Chiang2026} shows that stirrers might in fact contribute to the collisional cascade through erosive, non-disruptive collisions. Naturally, for this to occur, the stirrers' total mass needs to be higher than that in the collisional cascade.

\begin{figure}
    \centering
    \includegraphics[width=\columnwidth]{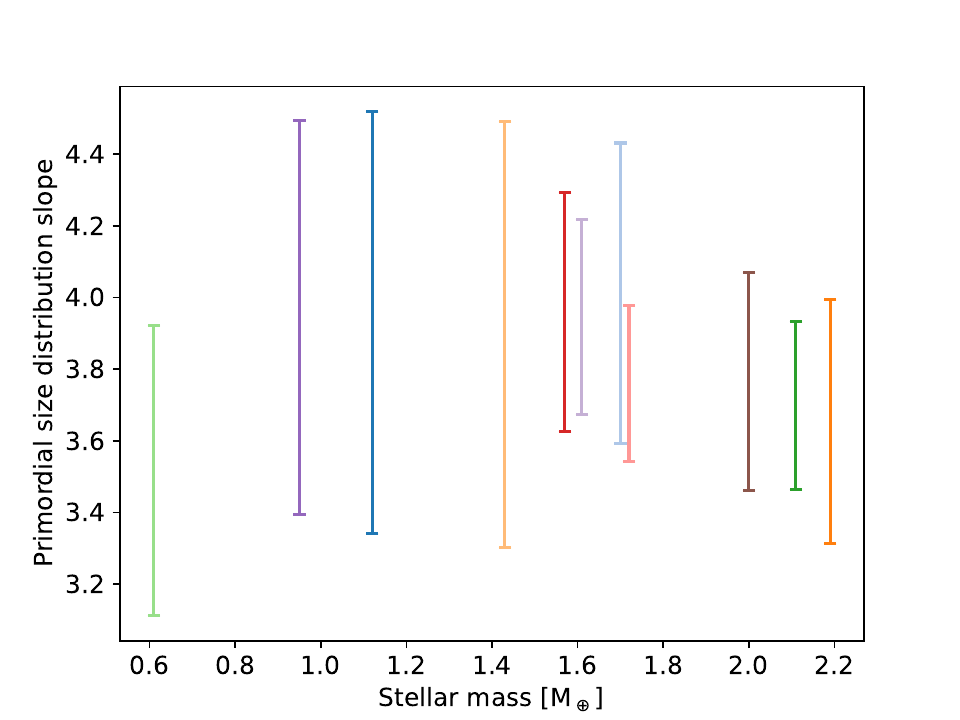}
    \caption{Primordial size distribution slope ranges for hypothetical power laws connecting constraints on planetesimals and constraints on stirrers. Upper limits are set by extrapolation of planetesimal sizes to sizes of largest possible stirrers, and vice versa for lower limits.}
    \label{fig:q_p_from_together}
\end{figure}

In our modelling, we did not assume that a particular distribution extends from planetesimal sizes up to stirrers' sizes. If we go on to assume a posteriori that these bodies are connected by a power law, we can extract the possible ranges for the slopes from Fig. \ref{fig:together}. We show these in Fig. \ref{fig:q_p_from_together} as a function of stellar mass. The minimum values are determined by connecting our solutions for planetesimal sizes for $q_{\rm p}=3$ with the smallest possible stirrers; the latter are mostly determined by setting an upper limit on the total mass in stirrers to be less than the expected solid mass in protoplanetary discs. The maximum values connect our $q_{\rm p}=4$ solutions to the largest possible stirrers. The maximum values of these inferred primordial slopes may exhibit a weak decreasing trend towards more massive stars (with the exception of the notable outlier HD~197481, see the discussion in Sect. \ref{sec:discussion_planetesimals}). This might be partly explained by the dependence of our minimum surface density of stirrers on stellar mass through Eqs.~(\ref{eq:sigma_min}) and (\ref{eq:h_hill}).

The connection between the planetesimals feeding the cascades and any hypothetical larger bodies is influenced by several planet formation processes. Planetesimal formation via the streaming instabilities and subsequent gravitational collapse is predicted to result in a fairly flat power-law distribution of planetesimals, with an exponential cut-off at larger sizes \citep{Johansen2015,Simon2016,Schafer2017}. Subsequent planetesimal-planetesimal collisions and accretion of pebbles flattens the cut-off \citep{Johansen2015}. Since extrapolating our collisional cascade planetesimals up to the sizes predicted by these studies leads to the disc mass problem \citep[see Sects. \ref{sec:m_pl} and \ref{sec:discussion_planetesimals}, and][]{Krivov2021}, we refrain from making more detailed comparisons to planet formation theory.

\subsection{Uncertainties}
In this section, we discuss uncertainties within the paradigm we adopted, in which planetesimals feed the steady-state production of the observed dust, while a population of bodies embedded within the discs gravitationally stirs the discs. Further below we discuss alternatives to this paradigm.

\subsubsection{Observational uncertainties} \label{sec:obs_uncertainties}
First and foremost, our results are likely affected by the selection of targets for the ARKS programme, and may not be representative of all debris discs. ARKS discs are some of the brightest and consequently also some of the youngest among the general population \citep{Marino2026a}.

The uncertainties in the ARKS disc aspect ratio fits estimated by the fitting methods are relatively low \citep{Zawadzki2026}.
However,
fits with different methods or assuming different vertical profile shapes often yield strikingly different aspect ratios. The extreme example is HD~197481. Assuming Gaussian vertical density distribution yields $h = 0.023$ and $h=0.026$ with non-parametric methods, which are close to previous estimates \citep[e.g.][]{Daley2019}. Assuming Lorentzian distribution yields $h=0.0028$ \citep[the value adopted in this work; see Fig. 2 in][]{Zawadzki2026}. A factor of ten difference in $h$ translates to an order-of-magnitude difference in the collisional disc mass and four orders of magnitude in $\Sigma_{\rm stir} M_{\rm stir}$ of stirrers. Besides, \citet{Zawadzki2026} assume that the disc aspect ratio is fixed throughout the disc, whereas at least in some discs it might vary with radius \citep{Matra2019,Han2025}.

For our collisional modelling, one key quantity is used implicitly: the dust surface density. We estimated it using the dust masses, radii, and widths from simple Gaussian modelling of the ARKS observations from \citet{Marino2026a}, $\Sigma_{\rm dust} = M_{\rm dust}/(2\pi r \Delta r)$. A much more detailed analysis of ARKS discs' radial structures is presented by \citet{Han2026}, including deprojected, deconvolved radial profiles of dust surface density. Our more approximate estimates of dust surface densities agree with the peak dust surface densities in their Fig. 3 remarkably well, with only three exceptions: HD~32297, HD~61005, and HD~131835. These are the discs with narrow peaks imposed on radially extended discs. Even for these three discs, the peak surface density from detailed modelling is less than a factor of 2 or 3 larger than in our modelling. Therefore, our modelling should be accurate for the regions of peak density inside these discs. Radial variations in the surface density and other quantities could be used to uncover radial variations in properties of planetesimals and stirrers, especially for radially wide discs (e.g. HD~14055). However, we refrained from expanding our analysis in this direction because a key quantity -- the aspect ratio -- was constrained assuming it to be radially constant and possibly only representative of the peak-density region.

Another large source of uncertainty is the system ages. We assumed that the ages of the collisional cascades and the timescale of stirring are equal to the stellar ages compiled from literature by \citet{Matra2025} and \citet{Marino2026a}. Estimates of stellar ages can be affected by systematic errors \citep[e.g.][]{Guzman2025}. The values in literature vary for some discs, for example, from 15\,Myr \citep{Rodigas2014} to 30\,Myr \citep{Kalas2005} for HD~32297 (we assume 30\,Myr). Additionally, the relation of the stellar evolution `zero time' to the disc evolution `zero time' is highly uncertain. The zero time for pre-main sequence stellar evolution is also difficult to define: the transition from a collapsing protostar to a visible pre-main sequence star is a continuous, gradual process rather than a single, distinct event. Halving the system age would halve the collisional disc mass in planetesimals and double the constraint on stirrers, $\Sigma_{\rm stir} M_{\rm stir}$.

\subsubsection{Material properties}
Our collisional modelling relies on assumptions about the material strength, determined by the material composition. If planetesimals were icier, they would be weaker than assumed here \citep{Benz1999,Wyatt2002,Leinhardt2009}. This would result in smaller critical projectile sizes and shorter collisional timescales. This, in turn, would result in larger planetesimal sizes and larger disc masses, possibly above the expected protoplanetary solid masses.
Additionally, collisional damping of weaker bodies would be even less important.

Also determined by material composition is the dust opacity, which is used to infer the dust mass from the observed (sub)millimetre flux. We refer the reader to the discussion in \citet{Krivov2021}, who find that plausible variations in material composition would not drastically change inferred disc masses.

\subsubsection{Size distribution} \label{sec:discussion_size_distr}
Our collisional masses depend strongly on assumptions about the size distribution.
These enter the calculations through the extrapolation of dust mass to larger sizes, and also through the fragmentation timescale via the number of colliders.

The size distribution at millimetre sizes can be probed observationally through measurements of the spectral index at millimetre wavelengths. Typically, studies find a wide range of values, with samples ranging from 4 to 15 discs: $q = 3.36 - 3.50$ \citep{Ricci2012,Ricci2015}, $q = 2.84 - 3.64$ \citep{MacGregor2016}, $q = 3.1 - 3.3$ \citep{Marshall2017}, $q = 3.2 - 3.7$ \citep{Norfolk2021}.  
From a sample of 133 discs, \citet{Marshall2026} find the dust size distribution slope has a wide distribution with a peak around $q=3.5$. For the discs in our sample they mostly find slopes very close to $q=3.5$, with the exception of HD~197481, with $q=3.1$. 
Therefore, the size distribution slope at dust sizes has a large scatter in observed discs.
From the theoretical standpoint, we expect $q=3.66$ for basalt and constant velocity dispersion \citep{Benz1999}, as assumed in this work, or a steeper slope if collisional damping is important and velocity dispersion varies with size \citep{Pan2012}. 

The initial size distribution of planetesimals is much less certain. Planetesimal formation simulations give some predictions \citep[e.g.][]{Johansen2015, Simon2016, Schafer2017, Gerbig2023}, yet these are incongruent with our results -- the disc mass problem discussed above. Additionally, the above studies have mostly focused on planetesimal sizes from gravitational collapse following streaming instabilities. Planetesimals may continue to evolve inside protoplanetary discs through mergers and through pebble accretion, which can re-shape their size distribution \citep[e.g.][]{Schlichting2011,Schlichting2013,Johansen2015}.

The inferred planetesimal sizes and corresponding disc masses depend not only on the assumed primordial power-law slope, but also on the size of the smallest primordial planetesimals. This is because the initial projectile-to-target size ratio has a huge effect on the initial fragmentation timescale. With an extreme assumption of a monodisperse (very narrow) initial distribution we found planetesimal sizes of only 10\,cm to 10\,m. Coincidentally, these are exactly the sizes we do not expect to form through coagulation in protoplanetary discs, because of the well-known `metre barrier' \citep[][]{Armitage2010}, and which might not be produced in great numbers by streaming instabilities either \citep{Klahr2020,Gerbig2023}. On the one hand, this can be interpreted as a sign that the initial size distribution is not too narrow. On the other hand, our model does not take into account how early fragmentation and cratering/erosion might quickly produce smaller bodies, changing the size distribution early on.  
Time-dependent numerical collisional models are needed to explore this further.

\subsubsection{Velocity evolution} \label{sec:discussion_velocity_evolution}
In our collisional
modelling, we assumed that the average collision velocity of planetesimals is constant throughout the system lifetime. This is contrary to our assumption that discs are self-stirred during their lifetimes, 
implying that \citep{Ida1993}
\begin{equation} \label{eq:v_rel_t_vs}
    v_{\rm rel}(t) = v_{\rm rel}(t_{\rm age}) \left( \frac{t}{t_{\rm age}} \right)^{1/4}.
\end{equation}
The fragmentation rate of planetesimals would be much smaller at early times, meaning that the planetesimals at the top of the cascade should be smaller than what we found. However, the correction factor can be roughly estimated by re-writing our condition for largest planetesimals in the cascade ($s_{\rm pl}$), $R_{\rm frag}(v_{\rm rel}(t_{\rm age})) t_{\rm age} = 1$ as
\begin{equation} \label{eq:R_frag_cond}
    \int_0^{t_{\rm age}} R_{\rm frag}(v_{\rm rel}(t)) \textrm{d}t = 1,
\end{equation}
effectively averaging the fragmentation rate over the velocity history. Inserting Eq.~ (\ref{eq:v_rel_t_vs}) into Eq.~(\ref{eq:R_frag_cond}), we find that the `corrected' planetesimal size $s_{\rm pl}$ is only 10--35\% smaller than inferred assuming constant collision velocity (for $q_{\rm p}=3-4$). The corrected masses in planetesimals vary by the same factor. We note that we did not account for mass evolution over time because the mass of these largest possible planetesimals in the cascade remains close to their initial mass.

Furthermore, our assumed stirring rate is correct if the velocity dispersion of the stirrers is equal to that of the bodies being stirred, or smaller. 
This is true if the velocity of the stirrers and of the smaller bodies grows at the same rate, or the stirrers are cooled by dynamical friction. However, if collisional damping acting on the dust is efficient, it may reduce the excitation of the dust below that of the stirrers \citep{Pan2012}. Alternatively, there could be one or more stirrers with stochastically larger eccentricities and/or inclinations; in such a case, secular effects might also become important \citep[][]{Costa2024}.

Lastly, we might question if our assumptions about stirring and damping are self-consistent. Specifically, we assumed that collisional damping is inefficient at high velocities. Considering Eq. (\ref{eq:v_rel_t_vs}), it could be argued that if damping was efficient at some prior point in time, the dust velocity growth should have halted at a lower value. This apparent inconsistency is easily resolved if one considers that planetesimals' velocities would still have been steadily increasing. Ultimately, the `early' dust is replaced by more recently produced dust with higher velocities.

\subsection{Alternative stirring and damping mechanisms} \label{sec:alternative}
Our assumption of stirrers embedded in discs, formed early on and stirring the debris from an initially zero velocity dispersion, yields reasonable results for the sizes and numbers of the stirrers. This does not guarantee that this scenario is indeed occurring in debris discs.

For example, many of the ARKS discs have a non-Gaussian vertical density distribution \citep{Zawadzki2026}, which is inconsistent with standard scattering theory where eccentricity and inclinations are in equipartition in the dispersion-dominated regime \citep{Ida1992}. A companion ARKS paper shows that such non-Gaussian profiles can arise when scattering occurs in the dispersion-dominated regime when eccentricity and inclinations are \textit{not} in equipartition, and instead eccentricities of either the debris or of the stirrer(s) are initially higher than the inclinations due to some other process \citep{Chiang2026}.

Furthermore, the growth of the stirrers may have been prolonged or perhaps is even still ongoing \citep{Kenyon2002b, Kenyon2004a, Kenyon2008, Kenyon2010}. \citet{Chiang2026} estimate that mergers are likely occurring among the stirrers in at least some ARKS discs. In this scenario, the stirrers had less time to stir the discs than assumed in our work, and the time period available for the planetesimals to have collided is shorter. The stirrers ought to be larger and/or more numerous than inferred in our work, and the planetesimals ought to be smaller \citep{Krivov2021}.

An alternative scenario is that discs are stirred by external perturbers through secular perturbations. In this scenario, one must distinguish between the question of perturbers stirring the velocity dispersion up to the critical value required for the collisional cascade to form, and the question of perturbers stirring the vertical aspect ratios to their observed values. The former only requires particle eccentricities to be stirred \citep[e.g.][]{Mustill2009,Pearce2022,Sefilian2024}, the latter requires a planet on an orbit initially inclined relative to the disc \citep[e.g.][]{Brady2023,Poblete2023,Sefilian2025}. Stirring could also be `mixed', caused by both internal and external big bodies \citep{Munoz-Gutierrez2023,Marshall2026b}, in which case the constraints derived here for the big bodies are upper limits. Vertical thicknesses of debris discs might also not be a product of gravitational interactions with planetary system bodies at all, and instead could be inherited from the protoplanetary disc phase \citep{Walmswell2013}, or induced by stellar flybys early on \citep{Ida2000, Kalas2000}.

Another possibly important omission in our work is the effect of gas, which is present in four discs in our sample: HD~39060, HD~9672, HD~32297, and HD~131835. If this gas is CO-dominated, second-generation gas, its density is too low to affect millimetre-sized dust grains \citep{Olofsson2022}. If, on the other hand, these discs contain massive remnants of protoplanetary gas, that gas would affect the aspect ratios measured at (sub)millimetre wavelengths \citep{Youdin2007,Olofsson2022,Marino2022}.

\section{Conclusions} \label{sec:conclusions}
We performed simple, semi-analytic collisional and dynamical modelling of a sample of 11 discs from the ARKS survey, whose vertical thicknesses are well constrained. Our findings are as follows:
\begin{enumerate}
  \item The dust observed at millimetre wavelengths can be explained as products of steady-state collisional cascades with planetesimals ranging from a few metres to ten kilometres in size, with total masses contained in planetesimals of less than 200\,M$_\oplus$. Larger planetesimals are not excluded by this modelling, but their presence may increase the disc mass above the expected mass budget of a planetary system.
  \item The discs' vertical thicknesses can be explained by gravitational stirring in the dispersion-dominated regime by embedded large bodies, with collisional damping counter-balancing the stirring in only two systems. Stirrers' individual masses are always larger than those of the planetesimals, but their total mass need not be.
\end{enumerate}

These results are consistent with expected masses of planetary systems. However, this is not a proof that the assumed processes indeed occur in debris discs. We see four ways to distinguish between different scenarios in the future. First, the population statistics could be improved by increasing the number of discs with well resolved vertical structure. We note, for example, that the ARKS survey found evidence of aspect ratios increasing with system age \citep{Zawadzki2026}, while the preceding REASONS survey \citep{Matra2025} with lower resolution data did not find any. Second, multiwavelength observations can constrain both the size distribution slope and the vertical structure dependence on the dust grain size \citep{Vizgan2022,Han2026b}. Third, in wider discs, the aspect ratio might vary with radius \citep[][Luppe et al., in prep.]{Matra2019,Han2025} and this should be explored in a larger sample. Fourth, \citet{Zawadzki2026} found that the majority of ARKS discs appear more likely to have a Lorentzian vertical density distribution, and not the Gaussian as is commonly assumed (including in this work). Observationally, confirming this finding in more discs and with higher quality data is a key to understand the processes shaping vertical structure. As for theory, models should go beyond simple predictions of the average collision velocity and make predictions for the inclination distribution in debris discs \citep[as in e.g.][]{Jankovic2024, Sefilian2025, Chiang2026}.

\begin{acknowledgements}
We thank the referee for their detailed, constructive, and very useful comments. MRJ acknowledges funding provided by the Institute of Physics Belgrade, through the grant by the Ministry of Science, Technological Development, and Innovations of the Republic of Serbia. TDP was supported for this work by a UKRI Stephen Hawking Fellowship and a Warwick Prize Fellowship, the latter made possible by a generous philanthropic donation. AAS is supported by the Heising-Simons Foundation through a 51 Pegasi b Fellowship. EC is supported by the Simons Investigator program and the Miller Institute for Basic Research in Science, University of California, Berkeley. CdB acknowledges support from the Spanish Ministerio de Ciencia, Innovaci\'on y Universidades (MICIU) and the European Regional Development Fund (ERDF) under reference PID2023-153342NB-I00/10.13039/501100011033, from the Beatriz Galindo Senior Fellowship BG22/00166 funded by the MICIU, and the support from the Universidad de La Laguna (ULL) and the Consejer\'ia de Econom\'ia, Conocimiento y Empleo of the Gobierno de Canarias. SM acknowledges funding by the Royal Society through a Royal Society University Research Fellowship (URF-R1-221669) and the European Union through the FEED ERC project (grant number 101162711). Support for BZ was provided by The Brinson Foundation. AMH acknowledges support from the National Science Foundation under Grant No. AST-2307920. JPM acknowledges research support by the National Science and Technology Council of Taiwan under grant NSTC 112-2112-M-001-032-MY3. SMM acknowledges funding by the European Union through the E-BEANS ERC project (grant number 100117693), and by the Irish research Council (IRC) under grant number IRCLA- 2022-3788. Views and opinions expressed are however those of the author(s) only and do not necessarily reflect those of the European Union or the European Research Council Executive Agency. Neither the European Union nor the granting authority can be held responsible for them.
\end{acknowledgements}

\bibliographystyle{aa} 
\bibliography{bibliography} 

\appendix

\begin{table*}[!h]
\section{Data and symbol tables}
\centering
\caption{Input data to our models \citep{Marino2026a,Zawadzki2026}, followed by results on the sizes and the masses in planetesimals feeding the collisional cascades.}
\label{table:data}
\begin{tabular}{l c c c c c c c c c c c} 
\hline\hline
Name & $M_*$ & $L_*$ & $t_{\rm age}$ & $f$ & $r$ & $\Delta r$ & $h$ & $M_{\rm dust}$ & $v_{\rm rel}$ & $s_{\rm pl}(q_{\rm p}=3)$ & $M_{\rm pl}(q_{\rm p}=3)$ \\
 & [$M_\odot$] & [$L_\odot$] & [Myr] & [$\times 10^{-4}$] & [au] & [au] &  & [$M_\oplus$] & [km\,s$^{-1}$] & [km] & [$M_\oplus$] \\
\hline

HD~10647 & 1.1 & 1.6 & 1700 & 2.6 & 104 & 65 & 0.053 & 0.03 & 0.6 & 3.9 & 9.7 \\
HD~131835 & 1.7 & 8.9 & 16 & 23 & 77 & 65 & 0.022 & 0.45 & 0.3 & 0.6 & 23.3 \\
HD~14055 & 2.2 & 29.0 & 300 & 0.89 & 169 & 162 & 0.054 & 0.05 & 0.6 & 0.02 & 0.2 \\
HD~15115 & 1.4 & 3.7 & 45 & 4.7 & 81 & 14 & 0.022 & 0.07 & 0.3 & 1.0 & 5.6 \\
HD~161868 & 2.1 & 24.0 & 300 & 0.65 & 121 & 134 & 0.19 & 0.03 & 2.6 & 1.7 & 5.5 \\
HD~197481~(AU~Mic) & 0.6 & 0.1 & 23 & 3.9 & 34 & 13 & 0.0028 & 0.012 & 0.04 & 0.003 & 0.03 \\
HD~32297 & 1.6 & 7.0 & 30 & 61 & 116 & 37 & 0.009 & 0.87 & 0.1 & 0.3 & 17.1 \\
HD~39060 ($\beta$~Pic) & 1.7 & 8.7 & 23 & 25 & 104 & 97 & 0.112 & 0.12 & 1.5 & 0.4 & 4.6 \\
HD~61005 & 0.9 & 0.6 & 45 & 25 & 72 & 36 & 0.015 & 0.15 & 0.2 & 0.4 & 4.3 \\
HD~76582 & 1.6 & 9.9 & 1200 & 2 & 204 & 176 & 0.192 & 0.12 & 1.8 & 3.3 & 39.4 \\
HD~9672 (49~Ceti) & 2.0 & 15.0 & 45 & 7.2 & 135 & 158 & 0.046 & 0.21 & 0.6 & 0.02 & 0.9 \\

 \hline
\end{tabular}
\end{table*}

\begin{table*}[h!]
\centering
\caption{Key parameters.}
\label{table:params}
\begin{tabular}{l l} 
\hline\hline
Symbol & Parameter \\
\hline

$M_*$ & Stellar mass \\
$L_*$ & Stellar luminosity \\
$t_{\rm age}$ & Stellar age \\
$f$ & Disc fractional luminosity \\
$r$ & Disc central radius \\
$\Delta r$ & Disc radial width \\
$h$ & Disc aspect ratio \\
$M_{\rm dust}$ & Dust mass \\
$\sqrt{\langle e^2 \rangle}$ & Root mean square of particle orbital eccentricities \\
$\sqrt{\langle i^2 \rangle}$, $i_{\rm rms}$ & Root mean square of particle orbital inclinations \\
$v_{\rm rel}$ & Collision velocity \\
$s$ & Particle size (radius) \\
$Q_{\rm D}^*(s, v_{\rm rel})$ & Specific critical disruption energy \\
$2 Q_{\rm D}^*/v_{\rm rel}^2$ & Critical projectile-to-target mass ratio \\
$n(s)$ & Differential size distribution \\
$M(s)$ & Mass in unit log-bin of size around $s$ \\
$s_{\rm t}$ & Target size \\
$s_{\rm p}$ & Projectile size \\
$s_{\rm c}$ & Critical projectile size \\
$s_{\rm blowout}$ & Blowout size from radiation pressure \\
$s_{\rm pl}$ & Size of planetesimals feeding cascades \\
$q_{\rm s}$ & Size distribution slope in the strength regime \\
$q_{\rm g}$ & Size distribution slope in the gravity regime \\
$q_{\rm c}$ & Size distribution slope around $s_{\rm c}$ \\
$q_{\rm p}$ & Primordial size distribution slope \\
$t_{\rm frag}$ & Fragmentation timescale \\
$t_{\rm damp}$ & Collisional damping timescale \\
$\rho_{\rm s}$ & Bulk density of particles in the cascade \\
$\rho_{\rm stir}$ & Bulk density of stirrers \\
$M_{\rm stir}$ & Mass of individual stirrer \\
$\Sigma_{\rm stir}$ & Surface density of stirrers \\

\hline
\end{tabular}
\end{table*}
\FloatBarrier

\section{Additional figures}
\begin{figure}
    \centering
    \includegraphics[width=\columnwidth]{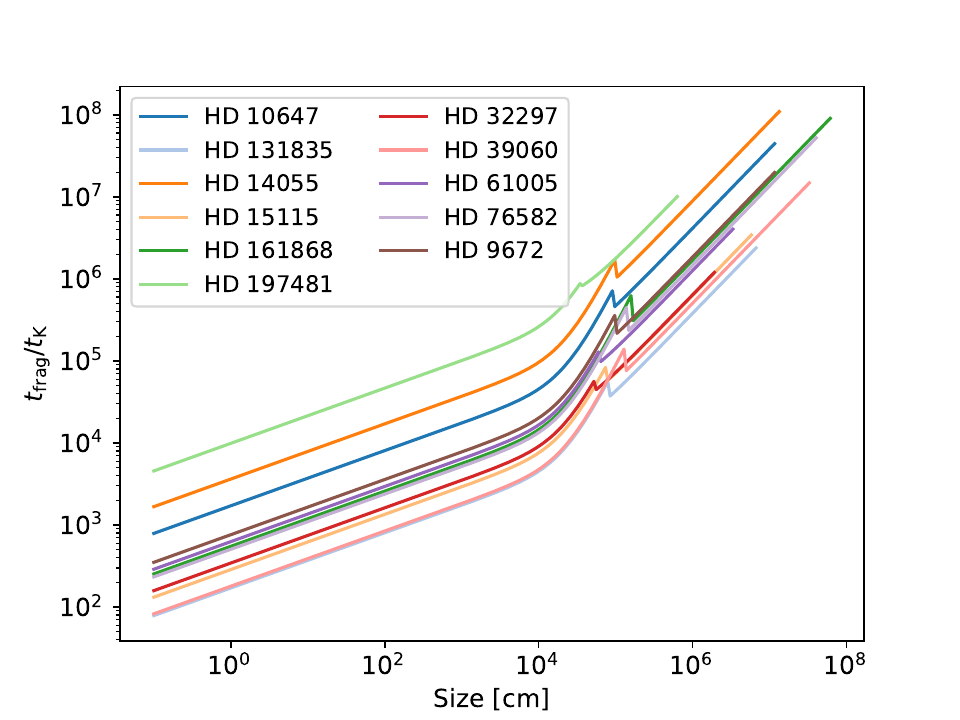}
    \caption{Ratio of the fragmentation timescale shown in the left panel of Fig. \ref{fig:t_frag} and the orbital period ($t_{\rm K}$) as function of particle size for each of the discs. Lines are cut off at the upper end at the sizes at which the critical disruption energy becomes too large and the bodies cannot be catastrophically disrupted except in collisions with larger bodies.}
    \label{fig:t_frag_t_k}
\end{figure}

Figure \ref{fig:t_frag_t_k} shows the ratio of the fragmentation timescale and the orbital period, as function of particle size. In all discs, for mm-sized dust and larger particles, catastrophic disruption collisions occur on average on the timescale of at least $\sim 100 \times$ the orbital timescale.

\section{Derivation of stirring constraints} \label{sec:derivations}
To derive constraints on properties of stirrer(s), we followed \citet{Matra2019}. We start from Eq. (4.1) from \citet{Ida1993}, which describes time evolution of the mean-square of particle eccentricities due to gravitational self-stirring, stirring from a protoplanet embedded in the disc and dynamical friction heating due to interaction with the protoplanet. As noted by \citet{Ida1993}, an analogous equation describes evolution of particle inclinations. Dynamical friction heating is at most as strong as viscous stirring \citep{Goldreich2004}, and we neglect it. We wish to consider the case of self-stirring, as well as the stirring by a protoplanet, as well as the case of stirring by a larger number of protoplanets or simply large bodies that are not planetesimals themselves. As shown by \citet{Ida1993}, these different scenarios can be considered in a unified manner. Neglecting a factor two difference between the rates in different scenarios, the stirring rate of the mean-square of inclinations is simply given by
\begin{equation} \label{eq:i_rate}
     \frac{\textrm{d} \langle i^2 \rangle}{\textrm{d} t} = \frac{\langle i^2 \rangle}{T_{\rm VS}},
\end{equation}
where $T_{\rm VS}$ is the viscous stirring timescale. From \citet{Ida1993}, in the dispersion-dominated regime the viscous timescale for the stirring of inclinations is given by their Eq. (4.2), replacing the numerical factor given therein for eccentricities with a factor $C_i = 2$ for inclinations,
\begin{equation} \label{eq:t_vs}
    T_{\rm VS} = \frac{1}{C_i} \left( \frac{\Omega_{\rm K}^2 r^3}{G M_{\rm stir}}  \right)^2 \frac{\langle i^2 \rangle^2}{\Sigma_{\rm stir}/M_{\rm stir} 
    \times r^2 \Omega_{\rm K}}.
\end{equation}
This expression holds both in the case of self-stirring and in the case of a single large stirrer, only in the latter case the surface density is an effective surface density \citep[see Eq. (4.8) in][]{Ida1993}. In the stirring rate and in the constraint derived below we do not make this distinction, and instead take it into account as an additional constraint on the minimum surface density of stirrers, as in Eq. (\ref{eq:sigma_min}), following \citet{Matra2019}.

Inserting $T_{\rm VS}$ from Eq. (\ref{eq:t_vs}) into Eq. (\ref{eq:i_rate}), we obtain
\begin{equation}
    \langle i^2 \rangle \textrm{d} \langle i^2 \rangle = \frac{C_i G^2 M_{\rm stir} \Sigma_{\rm stir}}{\Omega_{\rm K}^3 r^4} \textrm{d}t.
\end{equation}
Integrating from $t=0$ to $t=t_{\rm age}$ and from $\langle i^2 \rangle(t=0) = 0$ to the present-day value, assuming the spatial distribution and the sizes of the stirrers do not evolve, we obtain
\begin{equation}
    \frac{\langle i^2 \rangle^2}{2} = \frac{C_i G^2 M_{\rm stir} \Sigma_{\rm stir}}{\Omega_{\rm K}^3 r^4} t_{\rm age}.
\end{equation}
Inserting $\langle i^2 \rangle^2 = i_{\rm rms}^4$, $C_i$ and $\Omega_{\rm K}$, and re-arranging, this equation yields the constraint on $M_{\rm stir} \Sigma_{\rm stir}$ in Eq. (\ref{eq:m_sigma}).

The constraint on the stirrer(s) in the shear-dominated regime is derived analogously. From Eq. (4.7) of \citet{Ida1993}, neglecting a factor of $2^{2/3}$ difference between the self-stirring and stirring by a single large stirrer, the viscous timescale is given by
\begin{equation}
     T_{\rm VS, shear} = \frac{1}{C_i'} \left( \frac{\Omega_{\rm K}^2 r^3}{G M_{\rm stir}} \right)^{2/3} \frac{1}{\Sigma_{\rm stir}/M_{\rm stir} \times r^2 \Omega_{\rm K}} ,
\end{equation}
where $C_i'=3$. Inserting this into Eq. (\ref{eq:i_rate}), assuming $i_{\rm rms}(t=0)=i_0$, following the same steps as above and dropping the numerical pre-factor of an order of unity, we obtain Eq. (\ref{eq:m_sigma_shear}).

\end{document}